\documentclass[10pt]{article}

\usepackage{amsmath,amssymb,amsfonts,bm}
\usepackage{algorithmic}
\usepackage{graphicx}
\usepackage{dashrule}
\usepackage{textcomp}
\usepackage{xcolor}
\usepackage[preprint]{tmlr}
\usepackage{url}

\usepackage{silence}
\usepackage{natbib}
\setcitestyle{numbers}

\def\BibTeX{{\rm B\kern-.05em{\sc i\kern-.025em b}\kern-.08em
    T\kern-.1667em\lower.7ex\hbox{E}\kern-.125emX}}
\begin{document}

\title{LINOCS: Lookahead Inference of Networked Operators for Continuous Stability}

\author{\name Noga Mudrik \email nmudrik1@jhu.edu \\
      \addr Biomedical Engineering, Kavli NDI, CIS\\
      The Johns Hopkins University\\
      Baltimore, MD, 21218.
      \AND
      \name  Eva Yezerets \email  eyezere1@jhu.edu \\  
      \addr Biomedical Engineering, CIS\\
      The Johns Hopkins University\\
      Baltimore, MD, 21218
      \AND
      \name Yenho Chen \email   yenho@gatech.edu \\  
      \addr Department of Biomedical Engineering \\
      Georgia Institute of Technology \\
      Atlanta, GA 30332.
      \AND
      \name  Christopher J. Rozell \email  crozell@gatech.edu \\  
      \addr School of Electrical and Computer Engineering \\
      Georgia Institute of Technology  \\
      Atlanta, GA 30332.
      \AND
      \name Adam S. Charles \email adamsc@jhu.edu \\
      \addr Biomedical Engineering, Kavli NDI, CIS\\
      The Johns Hopkins University\\
      Baltimore, MD, 21218.}

\maketitle

\begin{abstract}


Identifying latent interactions within complex systems is key to unlocking deeper insights into their operational dynamics, including how their elements affect each other and contribute to the overall system behavior.
For instance,
in neuroscience, discovering neuron-to-neuron interactions is essential for understanding brain function; in ecology, recognizing the interactions among populations is key for understanding complex ecosystems. Such systems, often modeled as dynamical systems,  typically exhibit noisy high-dimensional and non-stationary temporal behavior that renders their identification challenging.
Existing dynamical system identification methods
often yield operators that accurately capture
short-term behavior  but  fail to predict long-term trends, suggesting an incomplete capture of the underlying process. 
Methods that consider extended forecasts (e.g., recurrent neural networks) lack explicit representations of element-wise interactions and require substantial training data, thereby failing to capture interpretable network operators.
Here we introduce \textbf{L}ookahead-driven \textbf{I}nference of \textbf{N}etworked\textbf{ O}perators for \textbf{C}ontinuous \textbf{S}tability (LINOCS), a robust learning procedure for identifying hidden dynamical interactions in noisy time-series data. LINOCS integrates several multi-step predictions with adaptive weights during training to recover dynamical operators that can yield accurate long-term predictions.
We demonstrate LINOCS' ability to recover the ground truth dynamical operators underlying synthetic time-series data for multiple dynamical systems models (including  linear, piece-wise linear, time-changing linear systems' decomposition, and regularized linear time-varying systems) as well as its capability to produce meaningful operators with robust reconstructions through various real-world examples.

\end{abstract}

\section{Introduction}

Uncovering the dynamics underlying high-dimensional time-series data is crucial for deciphering the fundamental principles that govern temporally evolving systems. 
This is apparent across significant scientific domains, including neuroscience (where neurons or ensembles interact over time~\citep{vaadia1995dynamics, d2019quantifying, mudrik2023sibblings}), 
immunology (where cells regulate immune responses~\citep{savill2002blast}), and ecology (where understanding population interactions yields insights into ecosystem dynamics~\citep{stein2013ecological}). 
Hence, scientific research necessitates the development of procedures adept at learning dynamic operators that can accurately capture the non-linear and non-stationary evolution of real systems. 


Existing approaches for dynamical systems identification, though, often rely on either ``black-box'' deep learning methods, which while powerful, yield uninterpretable representations, or on simple learning procedures that maximize reconstruction between consecutive samples, and thus
fail to accurately predict the system's behavior for long scales. 
Specifically, common dynamical systems identification models often rely on  optimizing the dynamics by minimizing the prediction error for each time point based on projecting the preceding one through the dynamics. Consequently, when using such procedures to learn the operators, post-learning long-term predictions of the system's values (by iteratively estimating the system's state  at the next time point) often result in undesired divergence away from the system.
This difficulty in long-term predictions, importantly,  implies that operators recovered by these models based on local cost functions may not capture the underlying system correctly.
The challenge in identifying such underlying operators, therefore, lies in the need to incorporate long-term predictions directly into the learning procedure, which can be especially challenging in cases where the dynamics are non-stationary, non-linear, or otherwise constrained in ways that reflect real-world system behavior. 

To address this challenge, we present a learning procedure that introduces Lookahead Inference of Networked Operators for Continuous Stability (LINOCS).
LINOCS bridges the gap between minimizing reconstruction costs based on single time-step projections (which often result in operators that quickly diverge in long-term forecasts), and optimizing multi-step training that relies on reconstructions from a past time point(which can lead to unstable predictions). LINOCS achieves this by integrating adaptively re-weighted multi-step reconstructions into the dynamics inference.
LINOCS also avoids relying on massive amounts of data (like RNNs, for example, require).
LINOCS' re-weighting progressively builds up the cost over training iterations, simultaneously considering several multi-step reconstruction terms for identifying operators that enable stable, long-term reconstruction post-training.
We demonstrate the effectiveness and  adaptability of LINOCS across a range of dynamical systems models, including linear, switched linear, decomposed systems, and smoothly  linear time-varying systems, achieving significantly improved accuracy in operator identification and long-term predictions.
Our contributions in this paper notably include:
\begin{itemize}
    \item We propose LINOCS, a novel learning procedure that incorporates re-weighting multi-step predictions into the cost for operator identification.
    \item We demonstrate that applying LINOCS improves the ability to recognize ground-truth operators.
    \item We show LINOCS' efficacy across a diverse range of dynamical systems, including linear, periodically linear, linear time-varying (LTV), and decomposed linear.
    \item Finally, we demonstrate LINOCS' ability to work on real-world brain recordings, resulting in better long-term reconstruction compared to baselines.
\end{itemize}

\section{Background and Terminology}
Consider a system with \(p\) interacting elements (e.g., neurons in the brain) whose time-changing state {\(\bm{X} \in \mathbb{R}^{p\times t}\)} evolves over discrete time points \(t = 1 \dots T\) as \(\bm{x}_{t+1} = g(\bm{x}_t,  \bm{b}_t, t)\), where $\bm{x}_t \in \mathcal{R}^p$ refers to the state at time $t$, \(\bm{b}_t \in \mathbb{R}^p\) represents an external input or driving force at time \(t\), and \(g\) is a function {\( {g: \mathbb{R}^p \times  \mathbb{R}^p \times \mathbb{Z} \rightarrow \mathbb{R}^p } \)}. 
For example, $\bm{x}_t$  can represent the time-evolving activity of $p$ recorded neurons over $T$ time points in neuroscience applications, or $\bm{x}_t$ can represent the activation levels of $p$ immune cells when modeling the immune system. 


In this paper we focus on linear, piece-wise linear,
and linear time-varying systems. 
Specifically, we limit our analysis to functions $g(\cdot)$ that can be written as:
\begin{align}
    \bm{x}_{t+1} = g(\bm{x}_t, \bm{b}_t, t) := \bm{A}_t \bm{x}_t + \bm{b}_t,
\end{align}
where $\bm{A}_t \in \mathbb{R}^{p\times p}$  represent the transition matrices  at each time $t$. 
Our focus on locally linear dynamics is supported by the fact  that even highly nonlinear functions can be well approximated over small time intervals using local linearization~\citep{khalil4control, sastry2013nonlinear}.
Importantly, this formulation's advantage lies in its retention of the ability to easily extract the system's pairwise interactions, including non-stationary changes in $\bm{A}_t$ and $\bm{b}_t$ over time. Specifically, any operator entry $[\bm{A}_t]_{i,j}$ for $i, j = 1 \dots p$ at every $t = 1 \dots T$ can be interpreted as the effect of element $j$ on element $i$ at time $t$.

In practice, however, robustly recovering operators that can  accurately describe the system's evolution in non-stationarity and non-linear settings,
faces numerous computational challenges.
Chiefly, if we adopt a naive approach to identify the operators as: $\widehat{\bm{A}}_t, \widehat{\bm{b}}_t = \arg \min_{\bm{A}_t, \bm{b}_t} \| \bm{x}_t - \bm{A}_t \bm{x}_t - \bm{b}_t \|_2^2$ for every $t = 1 \dots T$, this problem is statistically unidentifiable. Specifically, the problem has $p^2+p$ unknowns for each time point, but only $p$ equations.

One approach to improve inference in these settings is to introduce additional structure via a prior
over \( \bm{A}_t \) and \( \bm{b}_t \), 
that  constrains the solution space and is often grounded in application-driven assumptions. For instance, in many scientific settings, it is reasonable to assume that interactions change smoothly over time. Therefore, adding a temporal smoothness constraint on \( \bm{A} \) and \( \bm{b} \) (e.g., $\|\bm{A}_t - \bm{A}_{t-1}\|_F^2 $ and $\|\bm{b}_t - \bm{b}_{t-1} \|_2^2$) can be beneficial for both interpretability and accuracy. 
Additionally, the inclusion of such constraints can be crucial, particularly in noisy settings, to prevent overfitting. 
The addition of such constraints transforms the problem to:
\begin{equation}
     \widehat{\bm{A}}_t, \widehat{\bm{b}}_t = \arg \min_{ {\bm{A}}_t, {\bm{b}}_t } \|\bm{x}_{t+1} - \bm{A}_t \bm{x}_{t} - \bm{b}_t \|_2^2 + \mathcal{R}(\bm{A}_t, \bm{b}_t),
\end{equation}
where $\mathcal{R}(\bm{A}_t, \bm{b}_t)$ can represent regularization on the dynamics operators. 

While these solutions may offer good short-term predictions (i.e., predicting $\bm{x}_t$ from $\bm{x}_{t-1}$) with minor errors, they often struggle to fully reconstruct the dynamics over longer time-scales due to the build-up of
estimation errors over multiple predictions 
(e.g., if $\bm{x}_t|\bm{x}_{t-1}$ has a variance of $v_t$, then $\bm{x}_t|\bm{x}_{t-K}$ will have a variance of $\sum_{k=0}^{K-1} v_{t-k}$ due to the property of variance summation).
This issue of multiple potential solutions for the operators is further complicated by the practical consideration that we typically can only observe noisy observations of $\bm{x}_t$ (\(\widetilde{\bm{x}}_{t} = \bm{x}_t + \bm{\epsilon}_t\), where \(\bm{\epsilon} \in \mathbb{R}^{p\times T}\) represents some noise). Noisy observations further complicate the accurate identification of robust dynamics since if \(\widehat{\bm{A}}_t\) is obtained from \(\arg \min_{{\bm{A}}_t} \|\bm{x}_{t+1} - \bm{A}_t \widehat{\bm{x}}_t \|_2^2\), the distance between the real and the estimated operators \(\{ \|\bm{A}_t - \widehat{\bm{A}}_t \|_F^2 \}_{t=1}^T\) may increase with \(\bm{\epsilon}_t\).

As the accuracy of a dynamical system's fit to data is often evaluated based on its ability to accurately predict future values~\citep{tabar2019analysis}, the inability to capture long-term prediction suggests that the learned operators may have limited capacity to fully describe the system. 
Consequently, we define below three prediction styles for dynamical systems assessment that we will adhere to throughout this work:
\begin{itemize}
    \item \textbf{1-Step Prediction (\( \bm{x}_{t+1} | \bm{x}_{t} \)):} 1-Step Prediction  involves using the state at each time point \( t \) to estimate the state at the next time point (\( t+1 \)). 
   
    \item \textbf{Iterative Multi-Step Prediction (IMS)  of order \( K \in \mathcal{R} \) (\( \bm{x}_{t+k} | \bm{x}_{t}  \)):} IMS involves iteratively forecasting one-step ahead values and using these forecasts as inputs for further one-step ahead forecasts for $K$ times (i.e., $\forall k\in [0,K-1], \quad  \widehat{\bm{x}}_{t+k} | \widehat{\bm{x}}_{t+k-1}$, where $\widehat{\bm{x}}_{t-1} := \widetilde{\bm{x}}_{t-1}$). Here, we will notate an IMS prediction of order $K$ by  $ \widehat{\bm{x}}_{t}^K$.
    We chose to name this term IMS as to be consistent with the literature~\citep{chevillon2007direct}.

    \item \textbf{Full Lookahead Prediction (\( \bm{x}_{k}|\bm{x}_0\)):} 
This method enhances IMS by forecasting the state at each time point \(\bm{x}_t\) starting from the initial observations (\(\widetilde{\bm{x}}_0\)). It achieves this by sequentially applying transition matrices to the estimation from the previous time point \(\widehat{\bm{x}}_{t}|\widehat{\bm{x}}_{t-1}\), starting from \( \widetilde{\bm{x}}_{0}\), resulting in:
\(
\widehat{\bm{x}}_{t} = \widehat{\bm{A}}_{t-1}\dots\widehat{\bm{A}}_{0}\widetilde{\bm{x}}_{0} \quad \forall t=1 \dots T.
\)
\\
(Note: the formula above is presented without \(\bm{b}_t\) for simplicity, though they may be included).

\end{itemize}
Importantly, IMS Prediction and Full Lookahead Prediction  often result in instability due to the accumulation of errors in the sequential reconstructing process (Fig.~\ref{fig:introFig}A).

\vspace{5pt}
\subsection{Specific models considered in this work:}\label{sec:example_models}

Of particular interest in this work is improving the model fit of a core set of linear dynamical systems with different temporal constraints on the system evolution: 1) Time-Invariant Linear Dynamical Systems (LDS), 2) Switched Linear Dynamical Systems (SLDS), 
 3) decomposed Linear Dynamical Systems (dLDS), and 4) regularized Linear Time-Varying (LTV) Dynamical Systems. 

\textbf{Time-Invariant Linear Dynamical Systems (LDS).}
In linear systems analysis, the evolution of a general state $\bm{X}$ over $T+1$ time points can be typically represented as $\bm{x}_{t+1} = \bm{A} \bm{x}_t + \bm{b}$, where $\bm{A}\in \mathbb{R}^{p \times p}$ is the time-invariant dynamics matrix and $\bm{b} \in \mathbb{R}^{p\times 1}$ remain constant over time. 
One common method for determining $\bm{A}$ (and $\bm{b}$ if it is assumed that an unknown offset exists) involves a 1-step optimization approach that includes applying least squares across all time points. This entails solving 
\begin{align}
\widehat{\bm{A}} , \widehat{\bm{b}}= \arg \min_{\bm{A}, \bm{b}} \| \widetilde{\bm{X}}_{[:,1:T]} - \bm{A} \widetilde{\bm{X}}_{[:,0:T-1]} - [1]_{1\times T} \otimes \bm{b} \|_F^2 \label{eqn:step1linear}
,
\end{align}
where $\widetilde{\bm{X}}_{[:,1:T]}$ and $\widetilde{\bm{X}}_{[:,0:T-1]}$ 
represents the noisy observations of the state from the second time point ($t=1$) up to the last time point ($T$) and from the first time point ($t=0$) up to $T-1$, respectively, and  \([1]_{1\times T} \otimes \bm{b}\) represents the horizontal concatenation of the column vector \(\bm{b}\) horizontally \(T\) times. Here, $\bm{A}$ captures the average influence from $\bm{x}_{t-1}$ to $\bm{x}_{t}$ for all $t = 1 \dots (T+1)$.
This setting is advantageous in terms of ``network'' interpretability, however often cannot capture the complexity of real-world time-series which is non-linear and non-stationary.


\textbf{\textbf{S}witching \textbf{L}inear \textbf{D}ynamical \textbf{S}ystems (SLDS).
}
Switched Linear Dynamical Systems (SLDS)~\citep{Ackerson1970, Bar-Shalom1993, Hamilton1990, Ghahramani96switchingstate-space,murphy1998switching, NIPS2008_950a4152, linderman2017bayesian} 
aim to provide interpretable representations of dynamics by identifying linear operators that govern periods of linear behavior, with the system transitioning between these operators over time. 
Variations of SLDS include, e.g., recurrent SLDS (rSLDS), which introduces an additional dependency between discrete switches and the previous state's location in space~\citep{linderman2017bayesian}; and 
tree-structured recurrent SLDS, which extends rSLDS by incorporating a generalized stick-breaking procedure~\citep{nassar2018tree}.

While SLDS models often involve transitioning from an observed to a latent low-dimensional space, here we chose to focus on the case where switches occur within the observation space, essentially enforcing the transition to the latent space to be the identity operator. 
If we denote $\widetilde{\bm{X}} \in \mathbb{R}^{N \times T}$ as the noisy observations subjected to \textit{i.i.d} Gaussian noise, SLDS models the evolution of $\widetilde{\bm{x}}_t$ using a set of $J$ discrete states ($j = 1\dots J$), each state $j$ associated with its own linear dynamical system {$\bm{f}_j$}. These discrete states switch between them abruptly at certain time points following an HMM model. During each inter-switch period, 
if the system is in the $j$-th discrete state, SLDS models the evolution of the state linearly as $\bm{x}_t = \bm{f}_j \bm{x}_{t-1} + \bm{b}_j$, where $\bm{f}_j$ represents the linear transition matrix for the $j$-th discrete state and $\bm{b}_j$ denotes a constant offset term for that discrete state. 
SLDS can be trained by  an alternating set of steps between the dynamics learning and the HMM update of the operators.
SLDS tackles the crucial task of capturing non-stationarities while preserving interpretability, but it inherently lacks the capability to distinguish between multiple co-occurring processes or overlapping subsystems.

\textbf{\textbf{d}ecomposed \textbf{L}inear \textbf{D}ynamical \textbf{S}ystems (\textbf{dLDS}).
}
The Decomposed Linear Dynamical Systems (dLDS,~\cite{mudrik2024decomposed}) model relaxed the time-invariant or piecewise constant limitation of LDSs and SLDSs to support the discovery of co-occurring processes while maintaining interpretability.
Here, for simplicity, we focus on the case where the dynamics evolution is described directly in the observation space, while the full model presented in~\citep{mudrik2024decomposed} supports learning the dynamics within an identified latent state.
Specifically, dLDS models the dynamics evolution $\widetilde{\bm{x}}_t = \bm{A}_t \widetilde{\bm{x}}_{t-1}$
using a sparse time-changing decomposition of linear dynamical operators such that $\bm{A}_t = \left( \Sigma_{j=1}^J \bm{f}_j c_{jt} \right)$, resulting in 
$\widetilde{\bm{x}}_t = \left( \Sigma_{j=1}^J \bm{f}_j c_{jt} \right)  \widetilde{\bm{x}}_{t-1}.$
These dynamical operators ($\{\bm{f}_{j}\}_{j=1}^J$) are global, i.e., not time dependent, and hence are interpretable globally. However, their time-changing weights ($\bm{c}_t$) enable modeling non-stationary and 
non-linear dynamics~(Fig.~\ref{fig:different_systems} right). 
Notably, dLDS is trained through an Expectation-Maximization (EM) procedure where the global dynamics operators $\{\bm{f}_j\}_{j=1}^J$  and their time-changing coefficients $\{\bm{c}_{jt}\}_{t=1}^T$ are updated iteratively to maximize the posteriors as:
\begin{eqnarray}
    \{\widehat{\bm{c}_t}\}_{t=1}^T & = & \arg \max_{\{c\}} P(\{\bm{c}_t\}_t|\widetilde{\bm{X}}, \{\bm{f}_j\} ) \\
    \{\widehat{\bm{f}_j}\}_{j=1}^J & = & \arg \max_{\{\bm{f}_j\}} P(\{\bm{f}_j\}|\widetilde{\bm{X}}, \{\bm{c}_t\} ).
\end{eqnarray}

Interestingly, dLDS can also capture 
linear or switching behaviors described earlier, by fixing the dLDS coefficients over time (for linear behavior, Fig.~\ref{fig:different_systems} far left) or supporting abrupt change of coefficients in specific time points (for switching behaviors, Fig.~\ref{fig:different_systems} middle left).
As dLDS estimates the parameters for each time $t$ solely based on the values of the preceding state at time $t-1$, it does not address the issue of inaccurate long-term prediction due to accumulated deviations.

\textbf{Smooth or Sparse Linear Time-Varying Systems (LTV).}
In this paper, we refer to LTV systems that can be described by:
$\bm{x}_{t+1} = \bm{A}_t\bm{x}_t$ for all $t = 1\dots T$.
We further assume that a regularization $\mathcal{R}(\bm{A}_t)$ may be applied to the operators  $\{\bm{A}_t\}_{t=1}^T$. This regularization can be inspired by the application (e.g., smoothness of operators over time, $\|\bm{A}_t - \bm{A}_{t-1}\|_F^2 < \epsilon_2$ or operator sparsity $\|\textrm{vec}(\bm{A}_{t})\|_0 < \epsilon_1$) and mitigates the ill-posed nature of finding $\bm{A}_t$ separately for each time point.

\vspace{5pt}
\subsubsection{Prior relevant approaches:}

Theoretical literature on long-term prediction instability traces back to \cite{cox1961prediction} and~\cite{ klein2019essay}, who respectively introduced exponential smoothing and direct estimation of distant future states. 
Subsequent studies, including~\cite{findley1983use,findley1985model,weiss1991multi,tiao1993robustness,lin1994forecasting,kang2003multi}, evaluated the effectiveness of dynamical-systems identification methods in yielding long-term predictions. Specifically, the first approach focuses on identifying dynamical operators by minimizing the reconstruction error of projecting the state from one time point to the next, which can subsequently be used for long-term predictions through IMS. The second method, direct forecasting, aims to identify a mapping function $\bm{F}_{kt}: \mathbb{R}^p \rightarrow \mathbb{R}^p$ that predicts states further into the future by $\bm{x}_{t+k} = \bm{F}_{kt} (\bm{x}_t)$, thereby skipping the explicit identification of intermediate dynamic operators.
While direct estimation naturally results in more stable long-term predictions compared to 1-step optimization, it fails to provide an interpretable ``network'' meaning to the operators 
(e.g., in neuroscience, understanding the brain's interactions entails discerning the time-changing fast transitions from $\bm{x}_t$ to $\bm{x}_{t+1}$). 
In fact, when~\cite{marcellino2006comparison} compared between iterated and direct estimations  using  macroeconomic data, they found that in contrast to previous assumptions, iterated forecasts outperform direct forecasts, especially when models can select long-lag specifications---raising questions about the appropriate approach for learning dynamical operators.

More recently,~\cite{venkatraman2015improving} proposed a general approach called DAD that reuses training data to build a no-regret
learner with multi-step prediction. However, DAD includes ``fixing'' and updating the model itself based on every step within the multi-step prediction. Moreover, the authors presented it as a general abstract  non-specific approach to consider without specific implementation details.
More imporantly, DAD does not discuss the possibility for priors over the operators (e.g. temporal smoothness) during the training nor did they consider the need to find operators that are not only expressive but also interpretable. 
Other learning procedures, including full forward and/or backward passes through e.g., Backpropagation Through Time (BPTT) as in Recurrent Neural Networks (RNNs), can partially handle long-term prediction instability as well as 
Lookahead extensions to RNNs, including~\cite{unni2023improving, xiao2023deep,yeung2019learning}, that leverage the Koopman operator for improved long-term prediction.
However, these methods often require extensive learning data and remain uninterpretable in the ``network'' sense, as non-linear state measurements in Koopman operators can obscure understanding of pairwise state interactions.

An additional approach to understanding dynamical systems involves identifying a sparse set of functions that jointly decompose the observations. For example, SINDy~(Sparse Identification of Nonlinear Dynamics,~\cite{brunton2016discovering}) utilizes a data-driven approach to discover governing fundamental equations from data using sparse regression.
Although SINDy and its extensions (e.g.,~\cite{kaheman2020sindy}) are promising for discovering governing data equations, such representation does not provide explicit insight into the time-changing interactions between the state elements. 

\section{LINOCS}
In LINOCS we aim to learn 
the unknown dynamic operators {$\{\widehat{\bm{A}}_t\}_{t=1}^T$} 
by integrating several multi-step predictions simultaneously into the inference procedure. 
This approach yields not only a more accurate full-lookahead post-learning reconstruction but also operators that are more closely aligned with the ground truth.
Particularly,  for every $t = 1\dots T$, LINOCS finds the most likely estimate of $\{\bm{A}_t, \bm{b}_t\}$ given $K$+1 ($K \in \mathbb{Z}_{\geq 0}$ hyper-parameter) multi-step reconstructions of orders $k=0...K$ with different weights $\{\bm{w}_k\}_{k=0}^K$: 
\begin{align}
    \widehat{\bm{A}}_t = \arg \min_{\bm{A}_t} \sum_{k=0}^K  w_k \| \widetilde{\bm{x}}_{t+1} - \bm{A}_{t} \widehat{\bm{x}}_{t}^k\|_F^2,
\end{align}
where $\widetilde{\bm{x}}_{t}, \widetilde{\bm{x}}_{t+1}$ are the observations at time $t$ and $t+1$, respectively. 
Let $\widehat{\bm{x}}_{t+1}^k$ ($k = 0 \dots K$) denote the recursive rule for predicting the state at time $t+1$, starting from the observations at $t-k$, where $\widehat{\bm{x}}_{t-k}$ is set to the observations at  time $t-k$, i.e., $\widehat{\bm{x}}_{t-k} := \widetilde{\bm{x}}_{t-k}$. Particularly, the $(k+1)$-th order multi-step prediction is defined by:
\begin{align}
    \widehat{\bm{x}}_{t+1}^k = 
    \widehat{\bm{A}}_{t} \widehat{\bm{A}}_{t-1} \widehat{\bm{A}}_{t-2} \dots \widehat{\bm{A}}_{t-k} \widetilde{\bm{x}}_{t-k}. 
\end{align}

The weights $\{\bm{w}_k\}_{k=0}^K$ associated with the orders $k=0 \dots K$ are dynamically adjusted throughout the inference process (Fig.~\ref{fig:introFig}B). This adjustment considers both the order number ($k$) and the current reconstruction error related to that order, $e_k$ (e.g., the $\ell_2$ norm,  $e_k = \|\widetilde{\bm{x}}_t - \widehat{\bm{x}}^k_t\|_2^2$). Unlike other multi-step methods (e.g.,~\cite{venkatraman2015improving}), LINOCS adapts the weights of the reconstruction orders to prioritize the minimization of large errors in lower orders before considering higher orders (Fig.~\ref{fig:introFig}B). Specifically, it gradually increases the weight of the best lookahead reconstruction until convergence conditions are satisfied.
In our implementations, the weights can be chosen from a list of built-in choices such as uniform, linearly decreasing, and exponentially decreasing weights. Additionally, our framework allows custom weight functions that suit their specific needs.
In the experiments presented in this paper, we concentrate on showcasing three specific options for the weights:

\begin{figure}[t!]
    \centering
    \includegraphics[width=0.99\textwidth]{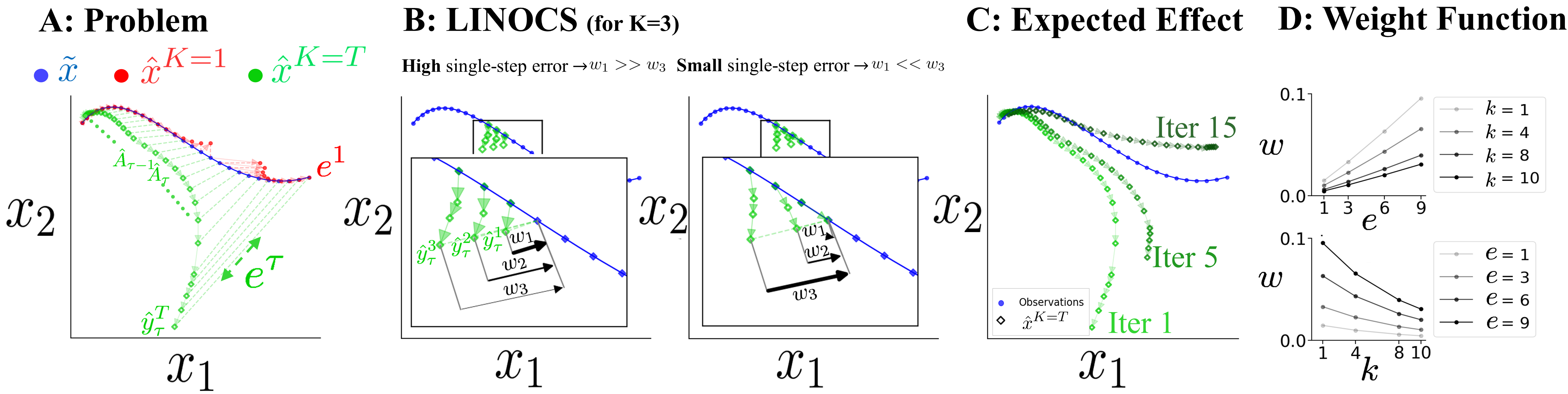}
    \caption{\textbf{Illustration of the problem and approach.} \textbf{A:}~Models that perform well on $1$-step prediction (i.e., prediction order $0$, in red) , often fail in higher orders reconstruction (e.g., order-$T$, in green).
    \textbf{B:}~LINOCS (e.g., for training order \( K=3 \)) integrates weighted multi-step reconstructions for all \( k = 0 \dots K \) orders. It adapts the weights of these reconstruction orders to prioritize minimizing large errors at lower orders before addressing higher orders. The system gradually increases the weight of the most effective lookahead reconstruction until convergence conditions are met.
    \textbf{C:}~LINOCS improves long-term reconstruction during training iterations.  
    \textbf{D:}~Weights of different lookahead training orders ($k$). $w_k: \mathbb{R}^2 \rightarrow \mathbb{R} $ is a function of the order $k$ and the $k$-th order reconstruction error $e$. Top: Illustration of an exemplary effect of $e$ on $w$ when fixing $k$. Bottom: Illustration of an exemplary effect of $k$ on $w$ when fixing $e$.
    }
    \label{fig:introFig}
\end{figure}

\begin{itemize}
    \item Adapting the weights to sequentially introduce higher multi-step reconstruction orders once the error for each preceding order falls below a designated threshold, while continuing to maintain the activation of lower orders. Specifically, in the initial iterations, only \( w_0 > 0 \), with all other weights \( w_j = 0 \) for \( j \in [1, K] \). As the error of each step’s reconstruction falls below this threshold, the subsequent weight \( w_{j+1} \) is activated. For instance, if \( w_0 \) is the last activated weight and the error for the first (1-step) reconstruction falls below the threshold, \( w_1 \) becomes active (\( w_1 > 0 \)), and this sequence of activation continues for higher-order weights as each subsequent step achieves the required accuracy.    
    
    \item Constant weights over iterations with an exponential decay over $k$, defined as $w_k = \exp^{-\sigma k}$ for some scalar $\sigma \in \mathbb{R}_{> 0}$.
    
    \item A weight function that considers both $k$ and $e_k$, exhibiting a monotonic decrease in $k$ and an increase in $e$, with $k$ decreasing faster than $e$ increases (Fig.~\ref{fig:introFig} D).
\end{itemize}

Importantly, throughout this paper, we distinguish between two concepts: training order and prediction order. We denote ``training order''  ($K_{\text{train}}$) as the maximum order considered \textit{during} LINOCS training. 
Throughout this work ``1-step'' optimization specifically refers to the use of the 1-step cost ($\|\bm{x}_t - \bm{A}_t\bm{x}_{t-1}|_2^2$), while excluding higher orders, \textit{during} training.
In contrast, prediction order refers to \textit{post-training} predictions that involve iteratively propagating the identified operators for $K_{\text{pred}}$ steps into the future.

Here, we demonstrate the contribution of LINOCS for accurate long-scale predictions in four types of systems: 1) time-invariant linear; 2) switching linear; 3) decomposed linear; and 4) LTV systems. 
Importantly, in our experiments, we assume that we observe the underlying system under additive \textit{i.i.d} Gaussian noise conditions, however LINOCS can be easily adjusted to other noise statistics.

\begin{figure}[t]
    \centering
    \includegraphics[width=0.99\textwidth]{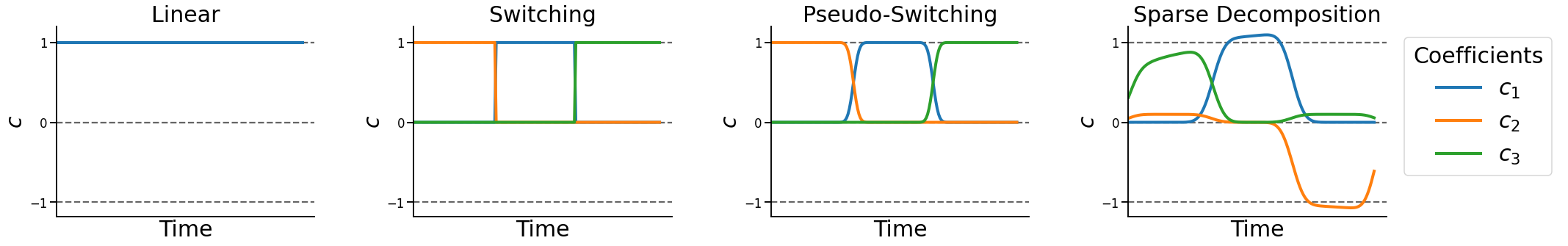}
    \caption{\textbf{Example time-varying LDS systems} \textbf{A:} The baseline linear time-invariant dynamical system will present constant dynamical operator constant over time. 
    \textbf{B:} Switching linear dynamical systems (SLDS) jump between different linear operators that are time-invariant between jumps.
    \textbf{C:} ``Pseudo-Switching'' dynamics is similar to SLDS with the inclusion of smoother transitions between periods of constant linear dynamics.  
    \textbf{D:} The decomposed linear dynamical systems (dLDS) model is a generalization of SLDS to sparse time-changing linear combinations of linear operators. dLDS can also model negative coefficients. 
    }
    \label{fig:different_systems}
\end{figure}

\begin{figure}[t!]
    \centering    \includegraphics[width=0.99\textwidth]{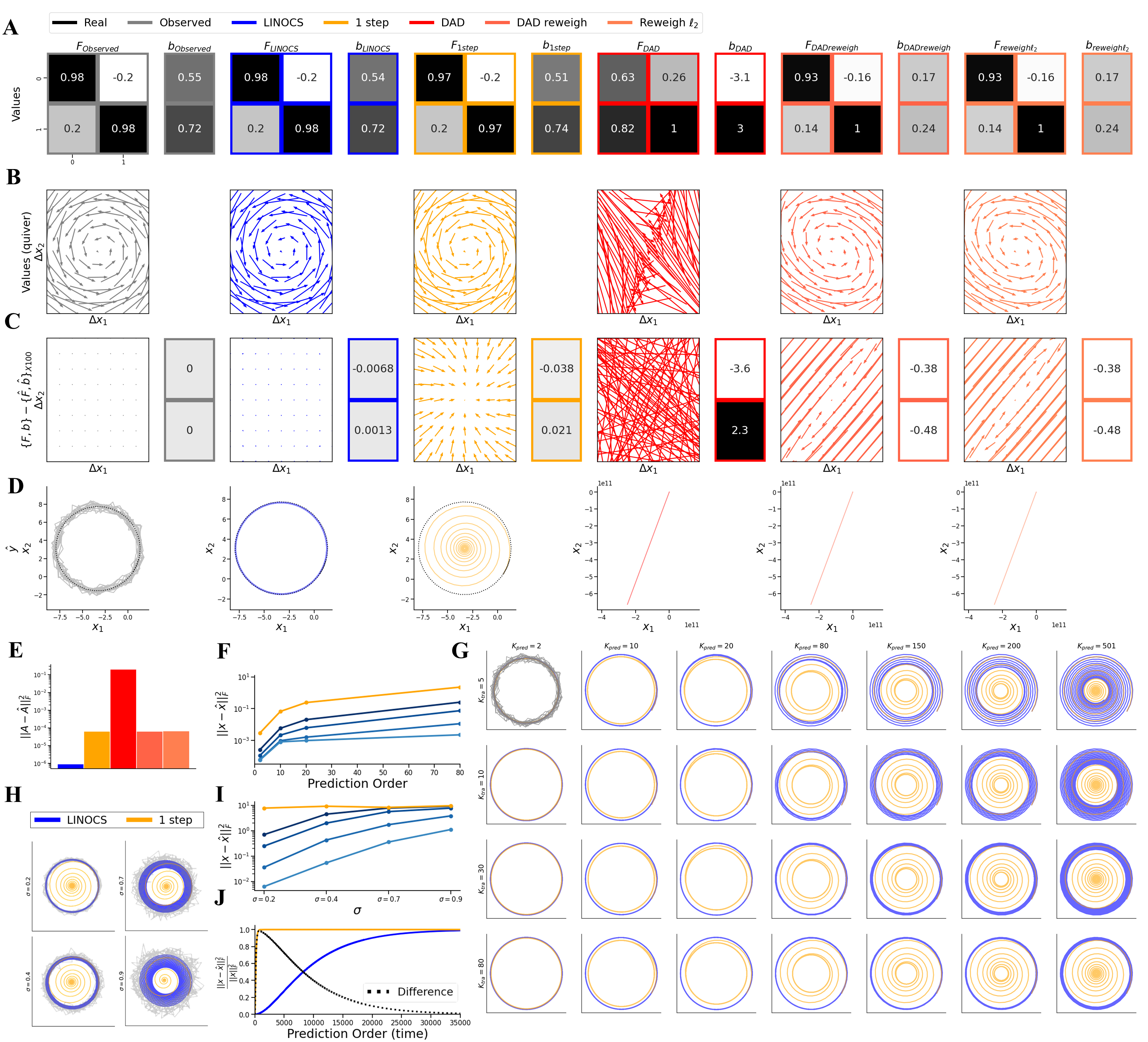}
    \caption{\textbf{Linear System Experiment.} 
    \textbf{A:} Real vs. identified operators and offsets. 
\textbf{B:} Quiver plots of real and identified operators. 
\textbf{C:} Highlighted differences in effects between real operators and inferred operators showing how small differences in dynamic operators gain prominence during lookahead reconstruction (calculation details in Section~\ref{sec:cal_details_linear_diff}).
    \textbf{D:} Full lookahead reconstructions (ground truth vs. baselines) show swift convergence to the origin for the 1-step optimization  (yellow) and divergence for DAD-based results (three most right subplots). 
    \textbf{E:} Frobenius norm of the differences between the ground truth operators ($\bm{A}$) and the identified operators ($\widehat{\bm{A}}$). 
    \textbf{F:} MSE under increasing prediction orders. For all orders, LINOCS achieves better (lower) MSE compared to 1-step optimization. 
    \textbf{G:} Full lookahead post-training predictions using operators identified by 1-step optimization (yellow) vs. the predictions using those identified by LINOCS (blue) under various training orders (rows) and prediction orders (columns).
   \textbf{H, I:} LINOCS reconstruction compared to 1-step optimization under increasing noise levels demonstrates its robustness.
    \textbf{J:} Propagating the identified operators until reaching a relative reconstruction error of $\sim 1$. LINOCS identifies operators that can accurately predict $\sim$ 35,000 time points, much higher than 1-step training that decay immediately. 
    }
    \label{fig:LINEAR_EXP}
\end{figure}

\subsection{LINOCS for Linear Dynamics}\label{sec:linear_section_exp}

We first present the LINOCS learning rule for the simplest case of time-invariant linear dynamical systems (TI-LDS). Let $\widetilde{\bm{X}} \in \mathbb{R}^{p \times T}$ be the observations of state $\bm{X}$, such that $\widetilde{\bm{X}} = \bm{X} + \eta$, with $\eta$ being an \textit{i.i.d} Gaussian noise ($\eta \sim \mathcal{N} (0, \sigma^2)$). In the TI-LDS case, $\bm{X}$ evolves linearly as $\bm{x}_{t+1} = \bm{A} \bm{x}_t + \bm{b}$ for all $t = 1 \dots T$, where $\bm{b} \in \mathbb{R}^{p\times 1}$ is a constant offset.

In this case, LINOCS estimates $\bm{A}$ and $\bm{b}$ by
$$\widehat{A}, \widehat{\bm{b}} = \arg \min_{\bm{A}, \bm{b}}  \sum_{k = 0}^{K} \bm{w}_{k} \|\widetilde{\bm{x}}_{t+1} - \bm{A}^{k+1} \widetilde{\bm{x}}_{t-k} -  \sum_{j=0}^k \bm{A}^{j} \bm{b} 
\|_2^2,$$ 
where $K$ is an hyperpameter that dictates the maximum  reconstruction order, and the set $\{\bm{w}_k\}_{k=0}^K$ can be either pre-defined or automatically adapted over training based on each order error.

\subsection{LINOCS for Switching Linear Dynamical Systems (SLDS)}\label{Sec:SLDS_APP}
For SLDS (Fig.~\ref{fig:different_systems}, middle-left), 
we integrate LINOCS into the SLDS operator inference stage 
(to infer $\{\bm{f}\}_{j=1}^J$, $\{\bm{b}\}_{j=1}^J$, see Sec.~\ref{sec:example_models}) using the SSM framework proposed by~\cite{Linderman_SSM_Bayesian_Learning_2020}. 
We maintain the existing SLDS approach to estimating switch times that delineate the boundaries of the linear periods between switches (i.e., for this stage, we have kept it as implemented by \cite{Linderman_SSM_Bayesian_Learning_2020}).
To find the operators within these identified linear periods, we integrate the learning rule for the TI-LDS case described above in Section~\ref{sec:linear_section_exp}.

\subsection{LINOCS for decomposed Linear Dynamical Systems (dLDS)}\label{Sec:dlds_exp}
For dLDS, as in SLDS, we incorporate LINOCS into the dynamical systems update step. 
Let $\widehat{\bm{x}}_{t+1}^k$ denote the $k$-th order reconstruction of $\bm{x}_{t+1}$, calculated by iteratively propagating the dLDS reconstruction $k+1$ times, starting from $\bm{x}_{t-k}$. 
Furthermore, let 
$\bm{x}_{t+1}\approx \Sigma_{j = 1}^J \widehat{c}_{jt} \widehat{\bm{f}}_j \bm{x}_{t}$, where $ \widehat{\bm{c}}_t$ represents the current estimate of the dLDS coefficients and $\{  \widehat{\bm{f}}_j \}_{j=1}^J$ denotes the current estimate of the basis operators.
We can now write the $k$-th order reconstruction ($\widehat{\bm{x}}_{t+1}^k$) as
\[
\widehat{\bm{x}}_{t+1}^k \leftarrow  \left( \sum_{j = 1}^J \widehat{c}_{jt}  \widehat{\bm{f}}_j \right) \left(\sum_{j = 1}^J \widehat{c}_{j(t-1)}  \widehat{\bm{f}}_j \right) \cdots  \left(\sum_{j = 1}^J \widehat{c}_{j(t-k)}  \widehat{\bm{f}}_j \right)\widetilde{\bm{x}}_{t-k},
\]
as follows from the recursive update rule 
$
\widehat{\bm{x}}_t^k \leftarrow (\sum_{j = 1}^J \widehat{c}_{jt}  \widehat{\bm{f}}_j)\widehat{\bm{x}}_{(t-1)}^{k-1}, \textrm{ where } \widehat{\bm{x}}^0_{t-1} := \widetilde{\bm{x}}_{t-1}
$.

To effectively integrate LINOCS into dLDS, we incorporate multi-step predictions into the training procedure of dLDS itself. Specifically, in each iteration, we start with the update of the dynamics coefficients $\bm{c}_t$. 
Second, we define $\bm{F}_{x_t^k} \in \mathbb{R}^{p \times J}$ as the horizontal concatenation $$\bm{F}_{x_t^k}:= [\bm{f}_1 \bm{x}_t^k,\bm{f}_2\bm{x}_t^k ,  \dots , \bm{f}_J \bm{x}_t^k] \textrm{ for all } k = 0\dots K.$$

Next,  we define a new matrix $\widetilde{\bm{F}}_{x_t}^{K}$ that extends $\bm{F}_{x_t}$ to all $K+1$ reconstructions stacked vertically, resulting in $\widetilde{\bm{F}}_{x_t}^{K} \in \mathbb{R}^{(K+1)p \times J}$
\[
\widetilde{\bm{F}}_{x_t}^K = \begin{bmatrix}
w_0 \bm{F}_{x_{t}} \\
w_1 \bm{F}_{x_{t}^{1}} \\
\vdots \\
w_{K-1} \bm{F}_{{x_{t}^{K}}}
\end{bmatrix},
\]
where $w_{k}$ is the weight of the $k$-th multi-step order. This matrix can then be used to infer to coefficients ($\bm{c}_t$) while considering different reconstruction orders with varying weights. 

To mirror this concatenated matrix of dynamics that represents multiple time-steps, we further define a matching concatenated state vector ${(\widetilde{\bm{x}}_{t+1})}_{vert} \in \mathbb{R}^{p(K+1) \times 1}$ 
by ${(\widetilde{\bm{x}}_{t+1})}_{vert}  := w \otimes \widetilde{\bm{x}}_{t+1}$ where {${w = [w_0; w_1;\cdots;w_K] \in \mathbb{R}^{(K+1) \times 1}}$}.
I.e., ${(\widetilde{\bm{x}}_{t+1})}_{vert}$ is obtained by
vertically stacking $K+1$ times the observations $\widetilde{\bm{x}}_{t+1} \in \mathbb{R}^{p\times 1}$ at time $t+1$ weighted by their corresponding $w_k$ values (resulting in 
\(
({\widetilde{\bm{x}}_{t+1}})_{vert} = [w_0 \widetilde{\bm{x}}_{t+1}; w_1 \widetilde{\bm{x}}_{t+1}; \cdots; w_{K} \widetilde{\bm{x}}_{t+1}]
\)).


The coefficients, $\bm{c}_t$, are thus updated in every iteration by minimizing the squared $\ell_2$ norm
\begin{align}
    \widehat{\bm{c}_t} = \arg \min_{\bm{c}_t} \| {(\widetilde{\bm{x}}_{t+1})}_{vert} - \widetilde{\bm{F}}_{x_t}^K \bm{c}_t \|_2^2. \label{eqn:dlds_estimate_of_c}
\end{align}

Note that $[\widetilde{\bm{F}}_{x_t}^K \bm{c}_t]\in \mathbb{R}^{(K+1) p \times 1}$ produces a vector of estimates of $\bm{x}_{t+1}$ computed from all different $K+1$ past states. This way, the estimator in Equation~\eqref{eqn:dlds_estimate_of_c} seeks the $\bm{c}_t$ vector that best predicts $\bm{x}_{t+1}$ considering all $K+1$ lookaheads.

One additional modification we make (compared to the original learning of dLDS as presented by~\cite{mudrik2024decomposed}) is that rather than updating each $\bm{f}_j$ using gradient descent, we infer the dLDS' basis dynamics operators $\{\bm{f}_j\}_{j=1}^J$ by fully and directly minimizing the cost. 
Specifically, let ${\bm{F}_{all} :=
[1]_{1\times J} = [\bm{f}_1, \bm{f}_2, \dots , \bm{f}_J] \in \mathbb{R}^{p \times {pJ}},}$
be the concatenated matrix of all $\{ \bm{f}_j \}_{j=1}^J$. 

Also, let  ${(\bm{x}_c)}_t := ([1]_{J\times 1} \otimes \bm{x}_t) \circ (\bm{c}_t \otimes [1]_{p \times 1}) \in \mathbb{R}^{pJ \times 1}$, where $\otimes$ denoted the Kronecker product and
$\circ$ denotes element-wise multiplication, 
and let 
$\bm{X}_c \in \mathbb{R}^{pJ \times T}$  be the horizontal concatenation of all ${{(\bm{x}_c)}_t \textrm{ (for } t = 0 \dots T-1)}$. 

With these definitions, the dLDS operators  $\{ \bm{f}_j \}_{j=1}^J$ are updated by
\begin{align}
{\widehat{\bm{F}}_{all}} = \arg \min_{{\bm{F}}_{all}} {\| \widetilde{\bm{X}}_{:,2:} - \bm{F}_{all} \bm{({X_c})} \|_F^2},
\end{align}
with each $\{ \bm{f}_j \}_{j=1}^J$ being extracted from $\widehat{\bm{F}}_{all}$ and normalized to a Frobenius norm of $1$.

\subsection{LINOCS for regularized Linear Time-Varying Systems (LTV)}

Finally, we focus on the more general
case of regularized linear time varying systems that are not necessarily switched or decomposed. Particularly, we focus on two types of regularizations, 1) that the operators change smoothly over time, i.e., $\| \bm{A}_t - \bm{A}_{t-1} \|_F^2$ is small, and 2) that the operators are sparse, i.e., $\| \bm{A}_t\|_0$ is small. 

For the LTV case, we apply LINOCS to find the 
time-changing operators $\{\bm{A}_t \}_{t=1}^T$ by iteratively integrating multi-step reconstruction with the appropriate regularization.
The operators are initialized with a regularized 1-step optimization 
$$
\widehat{\bm{A}}_t = \arg \min_{\bm{A}_t} \|\widetilde{\bm{x}}_t - \bm{A}_t \widetilde{\bm{x}}_{t-1} \|_F^2 + \mathcal{R}(\bm{A}_t).
$$

We integrate LINOCS into the estimation process by iteratively updating the operator estimates one at a time. Specifically, during each round of updates, we loop over every time point $t = 0 \dots T$ and, holding all other operators $\{ \bm{A}_{\tau} \}_{\tau \neq t}$ fixed at their former estimates, and update $\bm{A}_t$ by
$$\widehat{{\bm{A}}}_{t} = \arg \min_{\bm{A}_t} \sum_{k=0}^{K} \left [ w_k \sum_{k_i=1}^{k}\|\widetilde{\bm{x}}_{t-k_i+k} -\widehat{\bm{A}}_{t-k_i+k-1} \widehat{\bm{A}}_{t-k_i+k-2}\cdots \bm{A}_t \cdots \widehat{\bm{A}}_{t-k_i+1}  \widehat{\bm{A}}_{t-k_i}\widetilde{\bm{x}}_{t-k_i} \|_2^2 \right ] + \mathcal{R}(\bm{A}_t) $$ 
where $k = 0 \dots K$ denotes the order of the reconstruction and $t-k_i$ denotes the starting point of the reconstruction. The weights $\bm{w}^{k}$ are set as in all other models. 


\section{Results}

To showcase LINOCS' ability to capture the dynamics in multiple models, we applied LINOCS to the above systems under diverse settings. 
The hyper-parameters used in each experiment are summarized in Section~\ref{sec:hyperprameters}.

\subsection{LINOCS more accurately identifies ground truth linear systems under noisy observations}

We first test LINOCS' ability to robustly learn time-invariant 
linear dynamical systems from noisy observations. We then simulate the dynamics $\bm{A} \in \mathbb{R}^{2\times 2}$ as a rotational transition operator and a random offset $\bm{b} \in \mathbb{R}^{2\times 1}$, where each $\bm{b}_i \sim \text{Uniform}{(0,1)}$.
We build the synthetic state $\bm{x}_t\in \mathbb{R}^{2\times 1}$ as $\bm{x}_t = \bm{A} \bm{x}_{t-1} + \bm{b}$ starting from random initial value $\bm{x}_t \in \textrm{Uniform}(0,1)^{2\times 1}$,  such that the noisy observations are $\widetilde{\bm{x}} = \bm{x} + \eta, \textrm{where } \eta\sim \mathcal{N}(0, 0.3^2)$
(Fig.~\ref{fig:LINEAR_EXP}A,B). 

We compare the learned operators using LINOCS with four baselines. First we compare to traditional 1-step optimization  (Eqn.~\eqref{eqn:step1linear}). 
We further compare the linear LINOCS to our implementations of the conceptual framework presented in DAD~\citep{venkatraman2015improving}, as it is the approach closest to LINOCS in terms of integrating multi-step predictions into  model training. Our implementation of DAD integrates expert and non-expert demonstrations for model training, inspired by the Dataset-Aggregation (DAgger) approach~\citep{ross2011reduction}.
Specifically we test three implementations of DAD. For each implementation we initialized the transition matrix (\(\bm{A}_{init}\)) and the offset (\(\bm{b}_{init}\)) using the optimal estimate from 1-step optimization.
We then train the model through 100 iterations where at each iteration, we used our latest estimates of \(\bm{A}\) and \(\bm{b}\) to perform full lookahead reconstruction, starting from time $t=0$. 
We then update our estimates of \(\bm{A}\) and \(\bm{b}\) using the optimal 1-step optimization, and tested all three options outlined below:

\begin{itemize}
    \item  \textit{DAD with full model update:} \\
At each iteration, we update  $\bm{A}$ and $\bm{b}$ based on the lookahead reconstruction of the state ($\widehat{\bm{x}}_t$) calculated based on the last operators estimate. 
 Namely, ${\{\widehat{\bm{A}}_{\textrm{iter} + 1}, \widehat{\bm{b}}_{\textrm{iter} + 1} \} = \arg\min_{\{{\bm{A}, \bm{b} \}}} \frac{1}{T} \sum_{t=0}^{T-1} \|\widetilde{\bm{x}}_{t+1} - (\bm{A} \widehat{\bm{x}}_t + \bm{b}) \|_2^2}$. 

    \item \textit{DAgger-inspired DAD (reweighed DAD):}\\
For reweighted DAD,  we estimate $\bm{A}$ and $\bm{b}$ at each iteration using  both the observations and the lookahead reconstruction from the last estimates of $\bm{A}$ and $\bm{b}$. In particular, let  $\left[\widehat{\bm{x}}_{t}, \widetilde{\bm{x}}_{t}\right] \in \mathbb{R}^{p \times 2}$ be a horizontal concatenation of the lookahead reconstruction and of the observations at time $t$. 
Then we iteratively solve:
${\{\widehat{\bm{A}}_{\textrm{iter} + 1}, \widehat{\bm{b}}_{\textrm{iter} + 1}\} = \arg\min_{\{{\bm{A}, \bm{b}}\}} \frac{1}{T} \sum_{t=0}^{T-1} \| \left[\widehat{\bm{x}}_{t+1}, \widetilde{\bm{x}}_{t+1}\right] - (\bm{A} \left[\widehat{\bm{x}}_t, \widetilde{\bm{x}}_{t}\right] + \bm{b}) \|_2^2}$.

    \item \textit{DAgger with $\ell_2$ constraint (reweighted DAD with $\ell_2$):}
        \\ 
This option is solved similarly to the reweighted DAD, with the addition of the Frobenius norm ($\|\cdot\|_F$) on the operators ($\bm{A}$) and on ($\bm{b}$) during training. 
\end{itemize}

We find that LINOCS identifies operators that yield accurate dynamics in long-time scale predictions (Fig.~\ref{fig:LINEAR_EXP}D). 
The other methods we tested, including 1-step optimization (Fig.~\ref{fig:LINEAR_EXP}C,E cyan)  and DAD-based implementations (Fig.~\ref{fig:LINEAR_EXP}D, reds), instead decay to zero (away from the real system), indicating a less accurate estimation of the dynamics.
The improved accuracy of the operators identified by LINOCS (Fig.~\ref{fig:LINEAR_EXP}E) becomes apparent when examining the effects of the operators' estimation errors (Fig.~\ref{fig:LINEAR_EXP}C). These errors are larger for the other methods, showcasing that those methods accumulate more errors over shorter time spans than LINOCS' does. 

We next investigated the effect of the training order in LINOCS on long-term reconstruction.
We trained LINOCS on the noisy observations with increasing training orders (e.g. 5, 10, 30, 80) and then tested the performance under multiple prediction orders (Fig.~\ref{fig:LINEAR_EXP}G).
Compared to the baselines tested, LINOCS exhibits increased performance even with very low training orders (e.g., 5), with higher orders resulting in almost perfect reconstruction (Fig.~\ref{fig:LINEAR_EXP}G bottom-right subplot). Additionally, exploring LINOCS's robustness to noise reveals that, unlike one-step reconstruction, LINOCS is robust even under very high levels of noise (Fig.~\ref{fig:LINEAR_EXP}G, H, blue). The resulting MSE compared to the ground truth dynamics is much lower in LINOCS, even under very high $\sigma$ noise levels (Fig.~\ref{fig:LINEAR_EXP}H blue vs. orange-red).

When examining the duration for which LINOCS remains robust without converging, we observe that our approach accurately predicts approximately 35,000 time points into the future before deviating from the real system and decaying to $0$---demonstrating stability over exceptionally long time scales (Fig.~\ref{fig:LINEAR_EXP}J).

Additionally, when examining LINOCS'  long-term prediction robustness to increasing noise levels introduced during training, we find it to be robust to even extreme Gaussian noise levels, $\sigma = 0.9$ (Fig.~\ref{fig:LINEAR_EXP}D, blue), in contrast to 1-step optimization (Fig.~\ref{fig:LINEAR_EXP} D, orange-red).

We further tested LINOCS on linear systems with structured noise (Fig.~\ref{fig:LINEAR_EXP_noise}) as well as on a simulation of 3-dimensional cylinder
(Fig.~\ref{fig:LINEAR_EXP_3d}), yielding similar results. For structured noise, we modeled the observation as $\widetilde{\bm{x}} = \bm{x} + \sigma \sin(\gamma t)$ with $\sigma = 0.5$ and $\gamma = 3$ for $t = 1 \dots 501$. Unlike other methods, LINOCS found operators that led to accurate long-term predictions
(Fig.~\ref{fig:LINEAR_EXP_noise} D).
Moreover, when examined under increasing training and prediction orders, we found that LINOCS is robust for long-term predictions, even for full lookahead reconstructions ($k_\textrm{pred} = 501$, Fig.~\ref{fig:LINEAR_EXP_noise} E,H).
When evaluating its robustness to increasing noise levels ($\sigma$), we found that even for very high noise levels ($\sigma = 0.9$), LINOCS achieved much more robust results than 1-step optimization (Fig.~\ref{fig:LINEAR_EXP_noise} F,G).
Additionally, when exploring how far into the future it enables robust reconstruction before converging, we found that it is capable of full lookahead for approximately 70,000 time points—a testament to its ability to find more robust operators that can independently describe the system (Fig.~\ref{fig:LINEAR_EXP_noise} I).

For the 3D cylinder case (Fig.~\ref{fig:LINEAR_EXP_3d}), with Gaussian noise ($\sigma = 0.4$), we similarly demonstrate that LINOCS recovers more accurate operators, leading to significantly more robust long-term predictions and enabling full recovery of the process (Fig.~\ref{fig:LINEAR_EXP_3d} A,B,C,D), both under increasing prediction orders (Fig.~\ref{fig:LINEAR_EXP_3d} E, G) and noise ($\sigma$) levels (Fig.~\ref{fig:LINEAR_EXP_3d} H), as well as exhibiting an impressive ability to reconstruct lookahead predictions (starting from $\bm{x}_0$) for very long periods (approximately 70,000 time points) before converging to similar error as 1-step does (Fig.~\ref{fig:LINEAR_EXP_3d} I). In contrast, 1-step optimization yields high-error within a few IMS prediction orders.

\subsection{LINOCS identifies accurate interactions in switching systems}
\begin{figure}[t]
    \centering
    \includegraphics[width=0.99\textwidth]{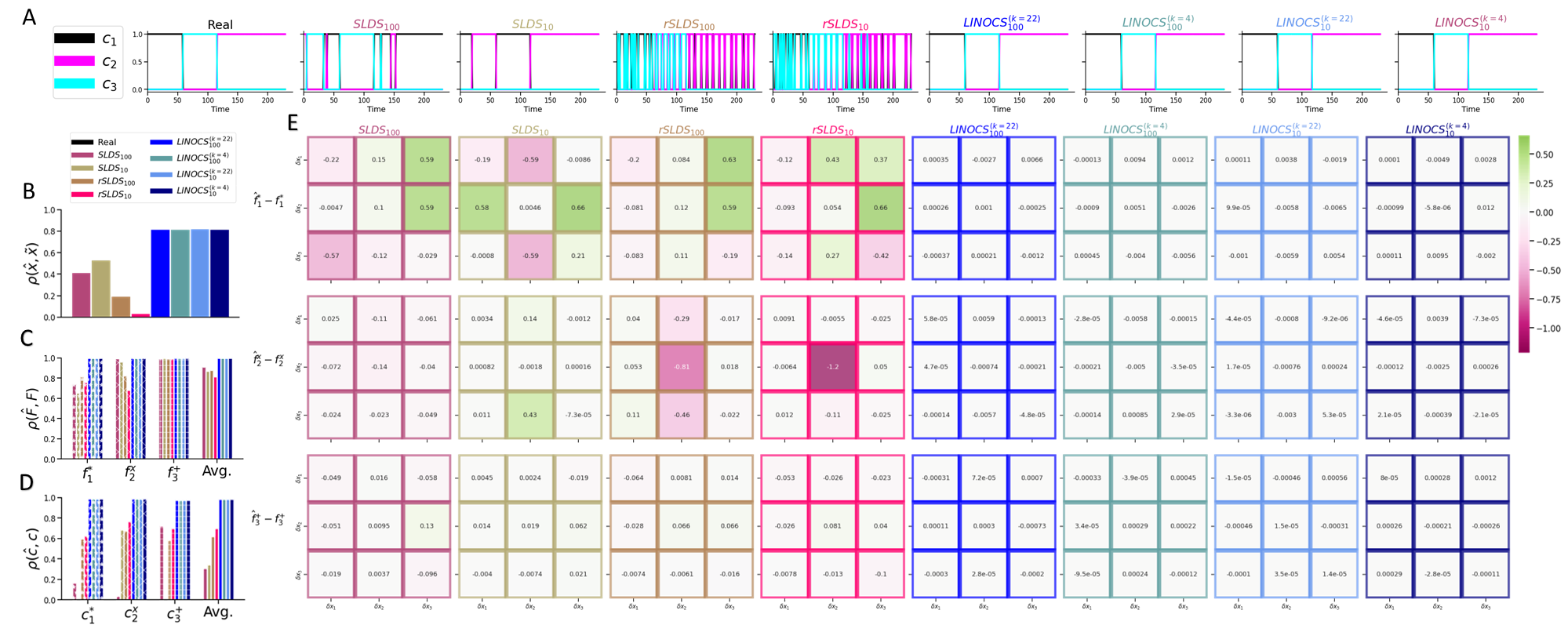}
    \caption{\textbf{Results on switching systems.} 
    \textbf{A:} Active discrete states for LINOCS (blue) compared to baselines, including SLDS and rSLDS with 10 or 100 training epochs. 
    \textbf{B:} Correlation between the ground truth dynamics ($\bm{x}$) and the full-lookahead reconstructed dynamics ($\widehat{\bm{x}}$). 
    \textbf{C:}  Correlation between the ground truth operators ($\bm{f}$) and the identified operators ($\widehat{\bm{f}}$). 
    \textbf{D:}  Correlation between the ground truth coefficients ($\bm{c}$) and the identified coefficients ($\widehat{\bm{c}}$). 
    \textbf{E:}  Difference between the ground-truth sub-dynamics ($\widehat{\bm{F}}$) and reconstructed basis dynamics by different models. LINOCS was able to achieve sub-dynamics that are much closer to the ground truth than the other baselines.  
    }
    \label{fig:SLDS_RESULTS}
\end{figure}

We next tested LINOCS-driven SLDS as detailed in Section~\ref{Sec:SLDS_APP} on simulated data comprising of $J = 3$ discrete states. The transition operators for each of the distinct states was set to a $3 \times 3$ rotational matrix oriented in a different direction. 
Additionally, the offset for each state 
($\bm{b}_j \in \mathbb{R}^{3 \times 1}$) 
was set to a random vector drawn from a uniform distribution between 0 and 1 (Fig.~\ref{fig:SLDS_synthetic_data}).

Notably,  since the method is invariant to the order of the operators, 
to compare the identified operators to the ground truth operators, we sorted the operators using   the ``linear sum assignment'' problem (SciPy's implementation, by ~\cite{crouse2016implementing}), with the cost function being the Frobenius norm between each pair of $\bm{f}$s (ground truth vs. estimated for each model). 
As baselines, we compare the results of LINOCS-augmented SLDS with standard SLDS and recurrent SLDS~\citep{linderman2016recurrent} with varying numbers of iterations.

When comparing LINOCS-SLDS to the baselines (Fig.~\ref{fig:SLDS_RESULTS}), LINOCS consistently outperformed the other approaches across multiple metrics including operator recovery (Fig.~\ref{fig:SLDS_RESULTS}C,E), switching times recovery (Fig.~\ref{fig:SLDS_RESULTS}A,D), and dynamics reconstruction (Fig.~\ref{fig:SLDS_RESULTS}B, Fig.~\ref{fig:switching_systems_supp}B).
In particular, LINOCS-SLDS accurately identified switching times, whereas classical SLDS and rSLDS tended to introduce additional redundant switches (Fig.~\ref{fig:SLDS_RESULTS}A).
Moreover, the discrepancies between the ground truth operators and those identified by LINOCS (Fig.~\ref{fig:SLDS_RESULTS}E, right-most  four columns) were substantially smaller than the differences observed with classical SLDS/rSLDS (Fig.~\ref{fig:SLDS_RESULTS} E, left columns),, as evidenced by the higher correlations between LINOCS' operators and the ground truth (Fig.~\ref{fig:SLDS_RESULTS} C). 
Furthermore, when examining the eigenvalues of the identified operators compared to the ground truth (Fig.~\ref{fig:slds_eigenvalues}), the eigenspectrum derived from the LINOCS-driven solver closely resembled the ground truth eigenspectrum more than the classical SLDS and rSLDS cases, highlighting the effectiveness of LINOCS in capturing the underlying dynamics.

\subsection{LINOCS finds dLDS operators that yield accurate dLDS lookahead predictions}
\begin{figure}[h]
    \centering
    \includegraphics[width=0.99\textwidth]{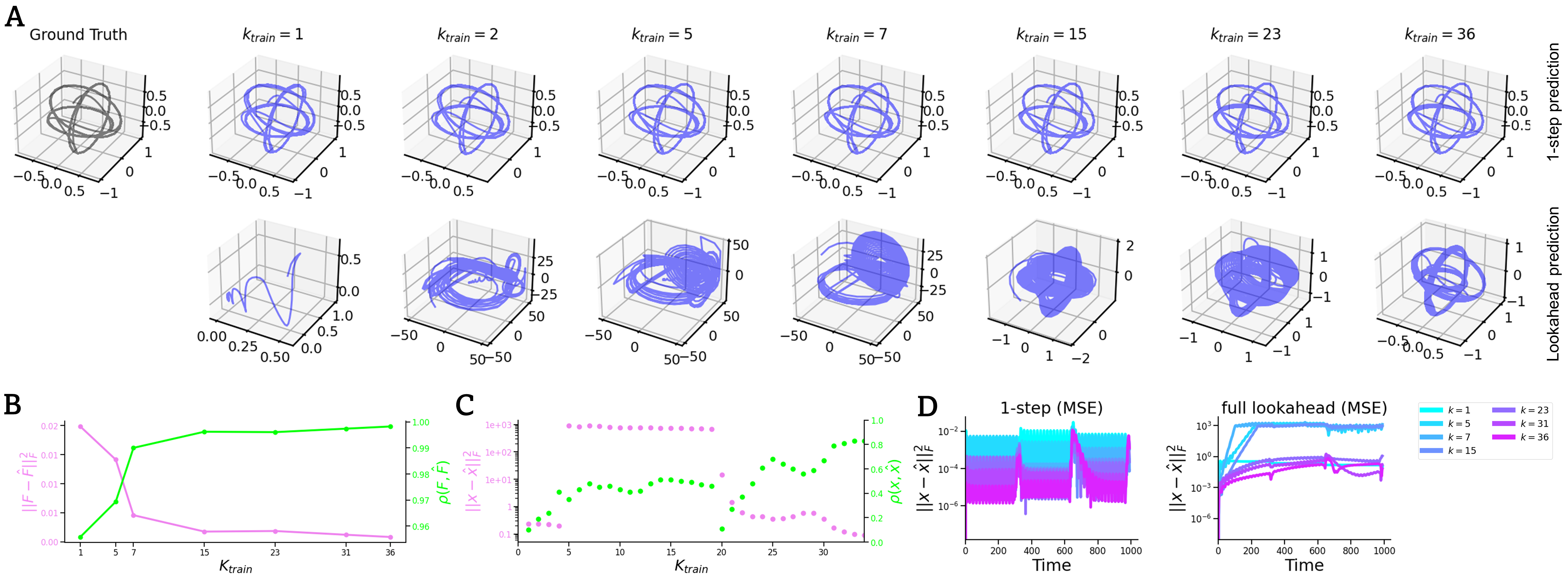}
    \caption{\textbf{Decomposed linear dynamical systems results.} 
    \textbf{A:} Ground truth dynamics compared to 1-step (top) and full lookahead (bottom) reconstructions for non-LINOCS dLDS ($K_{train} = 1$) and LINOCS-dLDS with different training orders.
     \textbf{B:} MSE (pink) and correlation (green) between ground truth operators  and the operators identified by LINOCS under different orders.  
      \textbf{C:} MSE (pink) and correlation (green) between ground truth dynamics and full lookahead reconstructions using the different LINOCS training orders.
      \textbf{D:} Local MSE for 1-step (left) and for full lookahead reconstruction (right) over the time points of the dynamics.  
    }
    \label{fig:dLDS_results}
\end{figure}

Next, we applied LINOCS to dLDS, as described in Section~\ref{Sec:dlds_exp}.
First we generated ground-truth data that represent a ``pseudo-switching'' (Fig.~\ref{fig:different_systems}) process---i.e. linear dynamics that switch more smoothly (in our case between $J=3$ systems) compared to SLDS where operators switch abruptly. This creates overlap periods where two dynamical systems are active at once as they trade off (Fig.~\ref{fig:dlds_ground_truth}). 
LINOCS-dLDS demonstrated significantly improved stability in full lookahead reconstruction compared to single-step dLDS (Fig.~\ref{fig:dLDS_results}). Notably, training with orders approximately greater than $35$ ($K_{\text{train}} > 35$) on our synthetic dataset (containing 1000 time points) resulted in high-quality full reconstruction (Fig.~\ref{fig:dLDS_results}A). Additionally, when comparing MSE and correlation of the time-evolving operator $\bm{F}_t = \sum_{j=1}^J c_{jt} \bm{f}_j$ to the ground truth, we observed a monotonic decrease in MSE with increasing maximal LINOCS training orders ($K_{\text{train}}$), while the correlation showed a monotonic increase (Fig.~\ref{fig:dLDS_results}B). Moreover, when comparing the largest eigenvalue of $\bm{F}_t$ over time between the ground truth operator and the identified operators, the high-order LINOCS-dLDS achieved a structure much closer to the ground truth (Fig.~\ref{fig:dlds_evals}).

Interestingly, although the 1-step reconstruction exhibited good performance even for non-LINOCS or low-order LINOCS-dLDS (Fig.~\ref{fig:dLDS_results}D left), the importance of LINOCS became more apparent in full lookahead reconstruction (Fig.~\ref{fig:dLDS_results}D right). This underscores the necessity of multi-step reconstructions for accurately estimating dynamical system dynamics, where errors might be obscured when only assessing the single-step reconstructions. It also highlights the importance of integrating multiple orders simultaneously during training.

\begin{figure}[ht]
    \centering
    \includegraphics[width=0.99\textwidth]{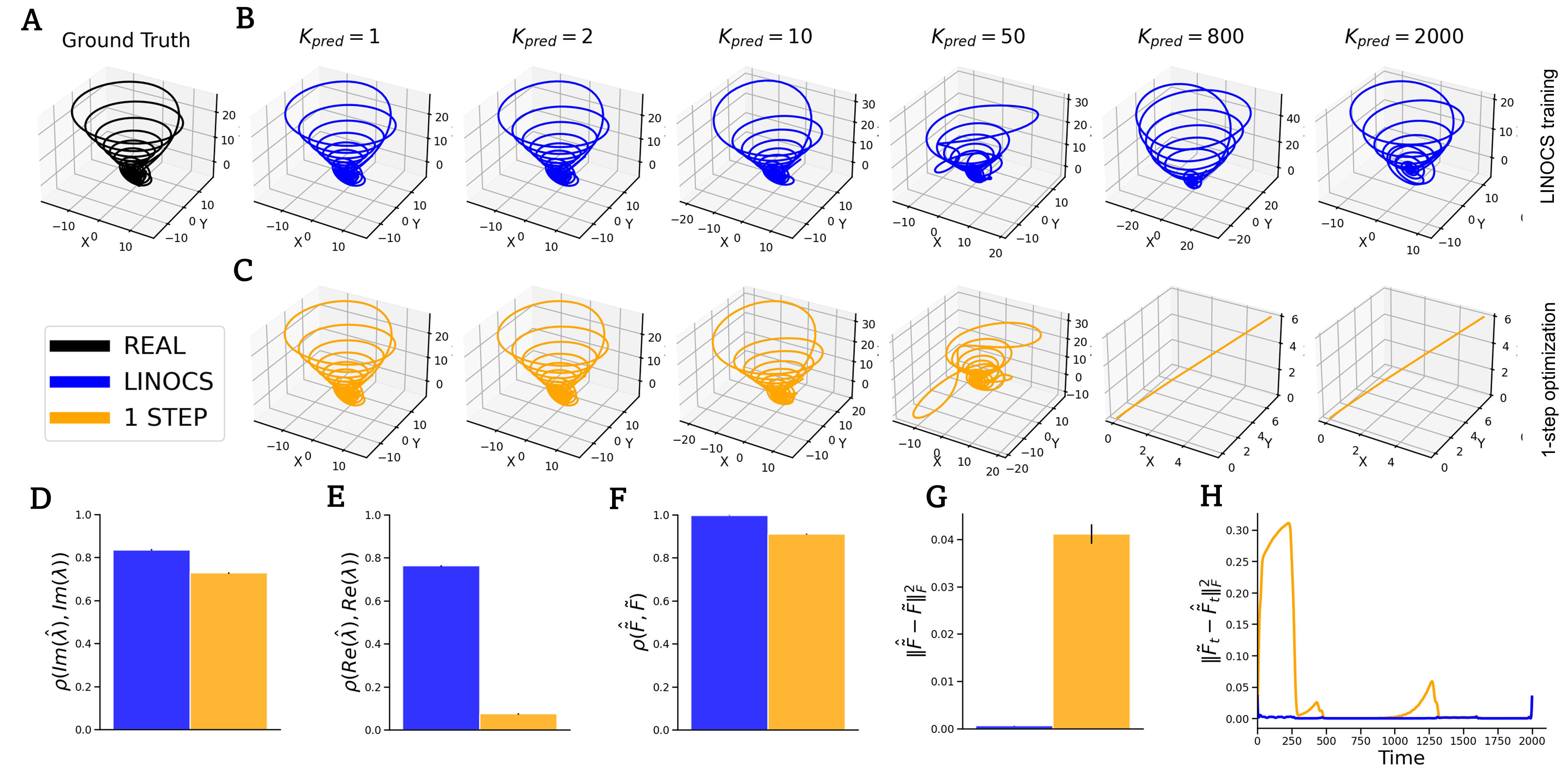}
    \caption{\textbf{Additional decomposed linear dynamical systems results.} 
    \textbf{A:} Ground truth dynamics compared to (\textbf{B}) LINOCS results and (\textbf{C}) 1-step optimization results, for increasing prediction orders. 
     \textbf{D,E:}~Correlation between the eigenvalues of the ground truth transition matrix (\({\bm{F}}_{t} = \sum_{j}^J c_{jt}\bm{f}_j\)) and the eigenvalues of the one identified by LINOCS.  \textbf{D:} imaginary part; \textbf{E:} real part. Results display the average correlation over time. Eigenvalues were matched using the ``linear sum assignment problem'' (Scipy's~\cite{crouse2016implementing}).      
      \textbf{F,G:}~Comparing the identified time-changing transition matrix  (${\bm{F}}_t$) identified by LINOCS vs. 1-step optimization in terms of correlation (\textbf{F}) and MSE (\textbf{G}). 
    \textbf{H:} Comparing the MSE of the identified ${\bm{F}}_t$ over time. 
    }
    \label{fig:dLDS_results_complex}
\end{figure}

We further extended our study to encompass more nuanced dLDS scenarios, exhibiting prolonged time scales and recurring patterns of identical active operators across distinct intervals (Fig.~\ref{fig:dlds_ground_truth_complex}). We found analogous enhancements of LINOCS over the traditional 1-step dLDS implementation. Specifically, LINOCS demonstrates robust accurate  long-term predictions, including full lookahead prediction (Fig.~\ref{fig:dLDS_results_complex}B), in contrast to 1-step optimization, which yield high lookahead error (Fig.~\ref{fig:dLDS_results_complex}C, last three subplots). Furthermore, also for this more complex example, upon comparing the identified time-varying transition operators ${\bm{F}}_{t} = \sum_{j}^J c_{jt}\bm{f}_j$ to the ground truth, LINOCS revealed operators with eigenvalues significantly more correlated with the real operators' evaluations (Fig.~\ref{fig:dLDS_results_complex}D, E) compared to the 1-step optimization results. Additionally, when comparing the operators themselves against the ground truth, those identified by LINOCS exhibited higher correlation and smaller MSE with the ground truth compared to these identified by 1-step  dLDS (Fig.~\ref{fig:dLDS_results_complex}F,G,H).

\subsection{LINOCS finds interactions that yield robust lookahead predictions in time-varying systems}
\begin{figure}[t]
    \centering
    \includegraphics[width=0.99\textwidth]{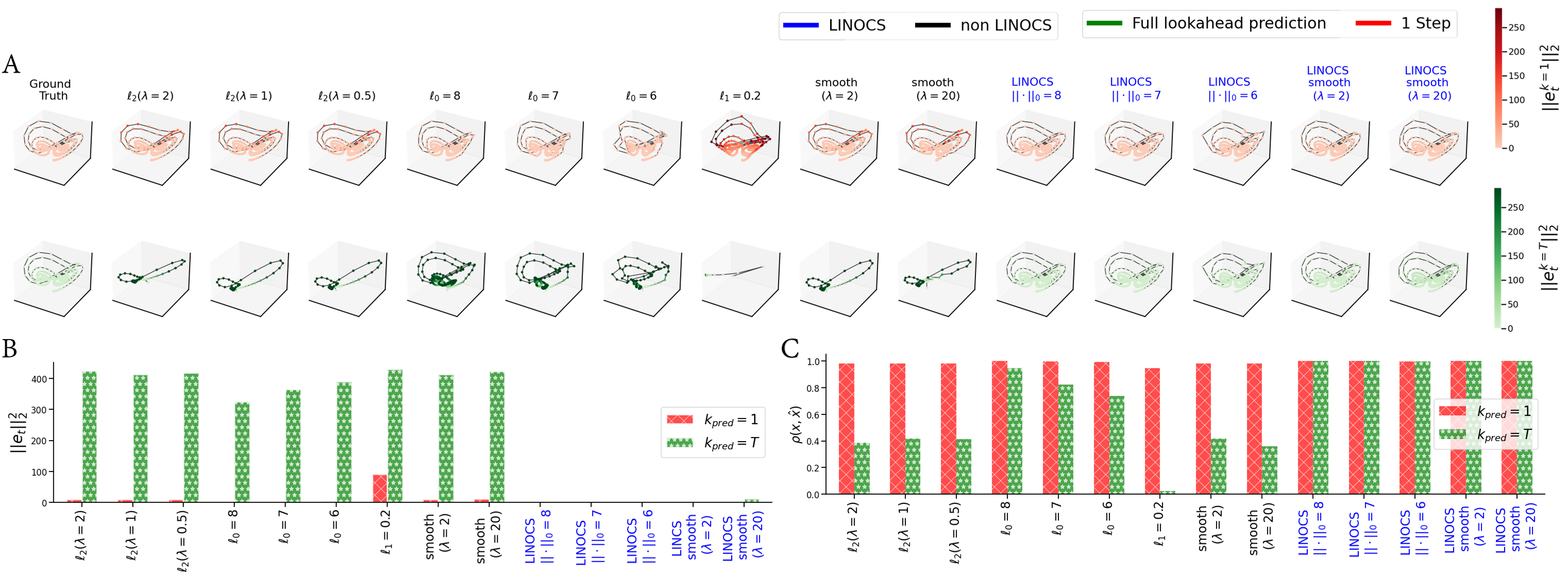}
\caption{\textbf{LTV approximation of the Lorenz attractor.}
    \textbf{A:} 1-step post-training prediction (top, pink-red) vs. full lookahead prediction (bottom, green) for different baselines, including LINOCS with various smoothness levels or $\ell_0$ regularization. Color indicates the local MSE.
    \textbf{B:} Squared $\ell_2$ of the error between the ground truth and 1-step prediction (red) vs. full lookahead predictions (starry green), for the different methods.
    \textbf{C:} Correlations between the ground truth and 1-step prediction (red) vs. full lookahead predictions (starry green), for the different methods.
}
\label{fig:LTV_RESULTS}
\end{figure}

To test the applicability of  LINOCS to more general LTV systems, we implemented LINOCS-LTV 
to capture the chaotic behavior of the Lorenz attractor (Sec.~\ref{sec:lorenz}) through a smoothly changing  LTV approximation
(Fig.~\ref{fig:LTV_RESULTS}).
We compared LINOCS-LTV with several other LTV solvers with varying constraints, including smoothness and sparsity. Unlike methods relying on 1-step optimization, LINOCS, despite similar regularization constraints, achieved superior full lookahead reconstruction (Fig.~\ref{fig:LTV_RESULTS}A bottom).

Also here, while different methods performed satisfactorily in the 1-step (post-training) prediction (Fig.~\ref{fig:LTV_RESULTS}A top, B red, C red), disparities emerged in higher-orders lookahead predictions where alternative methods failed. 
While all methods, including LINOCS, achieved commendable 1-step reconstruction, LINOCS demonstrated a markedly lower full lookahead error (Fig.~\ref{fig:LTV_RESULTS}B green, 5 most right bar pairs) and superior correlation with the ground truth (Fig.~\ref{fig:LTV_RESULTS}C green, five most right bar pairs). 

In addition,  we analyzed operators identified across various training iterations of LINOCS to assess their proficiency in achieving lookahead reconstruction (Fig.~\ref{fig:LTV_ITERATIONS}). 
For this analysis, we used the Lorenz attractor with 900 time points with intervals of 0.1/9 arbitrary units (a.u.), and applied a smoothness constraint with a weight of \(\lambda = 0.1\).
We observed that over training iterations, LINOCS adaptively influenced the predicted lookahead dynamics to gradually converge towards the ground truth dynamics  (Fig.~\ref{fig:LTV_ITERATIONS} C), with a monotonic decrease MSE (Fig.~\ref{fig:LTV_ITERATIONS}A, B). 

When analyzing which time points of the dynamics contributed to higher MSEs in the full post-training lookahead prediction, we noticed that early training iterations tended to produce higher full lookahead prediction errors at later time points of the dynamics (Fig.~\ref{fig:LTV_ITERATIONS}B, top right). However, over subsequent iterations, the effect of LINOCS managed to mitigate the accumulation of errors at these late time points (Fig.~\ref{fig:LTV_ITERATIONS}B, bottom  right).

\subsection{LINOCS finds robust interactions in real-world neural data}

Finally, we applied LINOCS to real-world  dataset described by~\cite{kyzar2024dataset}, which consists of high density electrode array of populations of single units in the human medial temporal and medial frontal lobes while subjects were engaged in a screening task.
We applied linear LINOCS, SLDS and dLDS-LINOCS, and LTV-LINOCS to a single recording session that includes recordings from five brain areas (amygdala left and right, cingulate cortex, hippocampus, pre-supplementary motor area). 
Dynamical systems models were trained on the firing rate data, which we inferred from the spike-sorted electrophysiology via a Gaussian kernel convolution. 
 


\begin{figure}[t]
    \centering
    \includegraphics[width=0.99\textwidth]{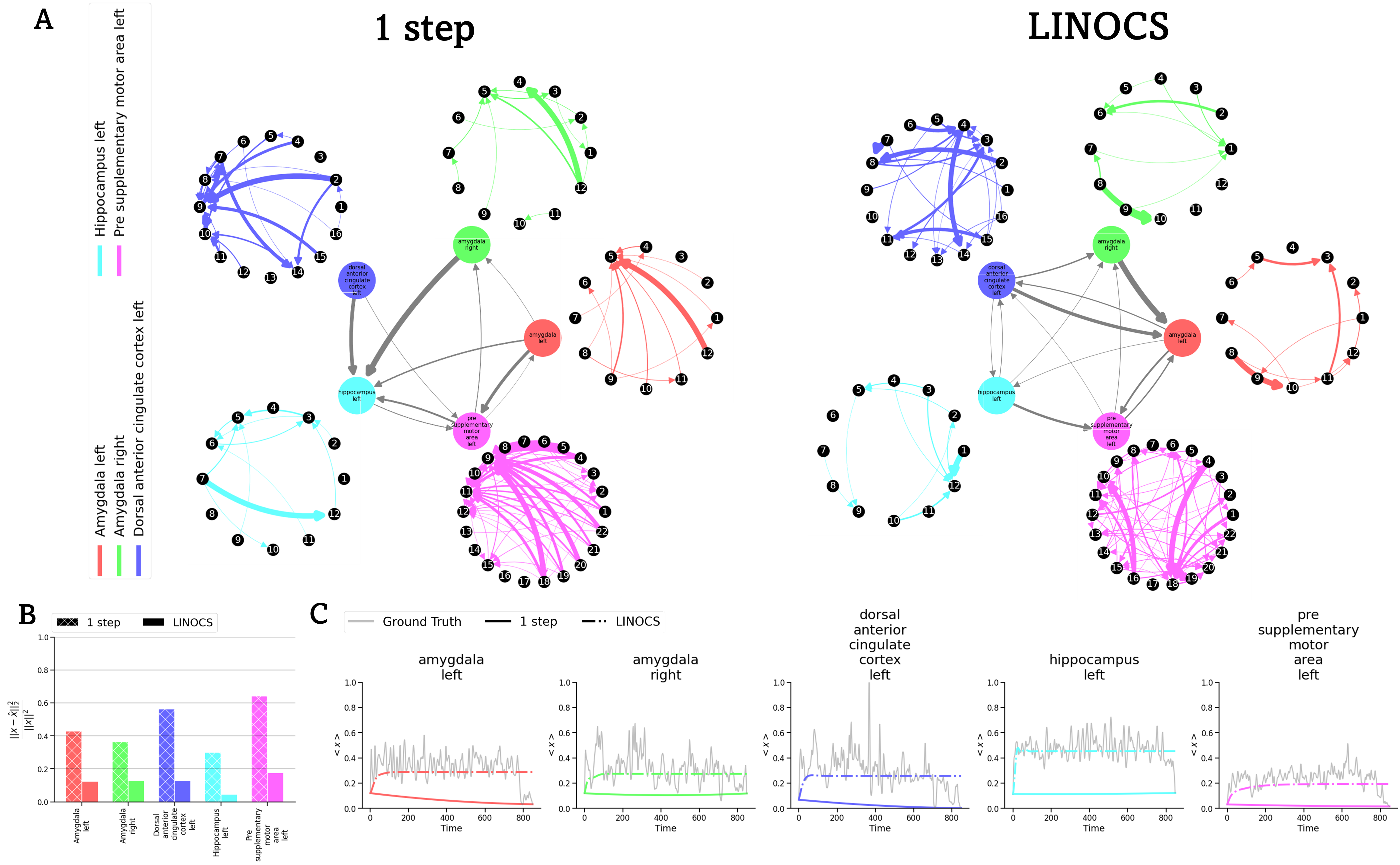}
   \caption{\textbf{Application of linear-LINOCS to multi-region neural recordings.} 
   \textbf{A:} The per-region and between-region mean field linear dynamics operators identified by 1-step linear optimization vs. LINOCS with a linear time-invariant system. 
   Each within-region network describes the linear operator $\widehat{\bm{A}}_{region}$ derived by applying LINOCS to the firing rate matrix constrained to include only the  neurons from that region. 
      \textbf{B:}~Full-lookahead reconstruction  error for 1-step linear optimization vs. LINOCS-linear approach.
   \textbf{C:} Ground truth mean activity per region compared to the lookahead prediction trace of the mean field activity by 1-step linear optimization vs. LINOCS-linear.
    }
    \label{fig:REAL_WORLD_LINEAR}
\end{figure}

We investigated several LINOCS models to showcase their distinct characteristics.
First, we examined the linear case for each brain area individually and explored the mean field interactions between areas (Fig.~\ref{fig:REAL_WORLD_LINEAR}). Importantly, while typical real-world brain dynamics 
are assumed to be non-linear and non-stationary, our aim in starting with the linear model was to demonstrate how LINOCS can identify the fundamental background neural interactions under linear assumptions and check how its identified interactions defer from these identified by the  1-step approach.
We first applied the linear LINOCS on the firing rate activity from all neurons within each region to identify between-region interactions.
We observed that LINOCS identified different linear interactions within areas compared to the 1-step approximation. Drawing from our conclusions based on synthetic data linear results, this suggests that LINOCS may better capture the linear approximation of brain activity than the common 1-step optimization (Fig.~\ref{fig:REAL_WORLD_LINEAR} A). 

 Then, we also applied the linear LINOCS on the mean activity of each region, and found that when examining the full lookahead reconstructions (Fig.~\ref{fig:REAL_WORLD_LINEAR} C) the 1-step optimization, in contrast to LINOCS, decayed to zero activity due to small accumulated deviations in operator values. In contrast,  LINOCS managed to maintain activity closer to the average values of the dynamics. However, due to linear enforcement, neither approach could capture fluctuations in dynamics.
 Moreover, the full lookahead reconstruction error for LINOCS-linear was overall much smaller compared to the classical 1-step (Fig.~\ref{fig:REAL_WORLD_LINEAR}B).

\begin{figure}[t]
    \centering
    \includegraphics[width=0.99\textwidth]{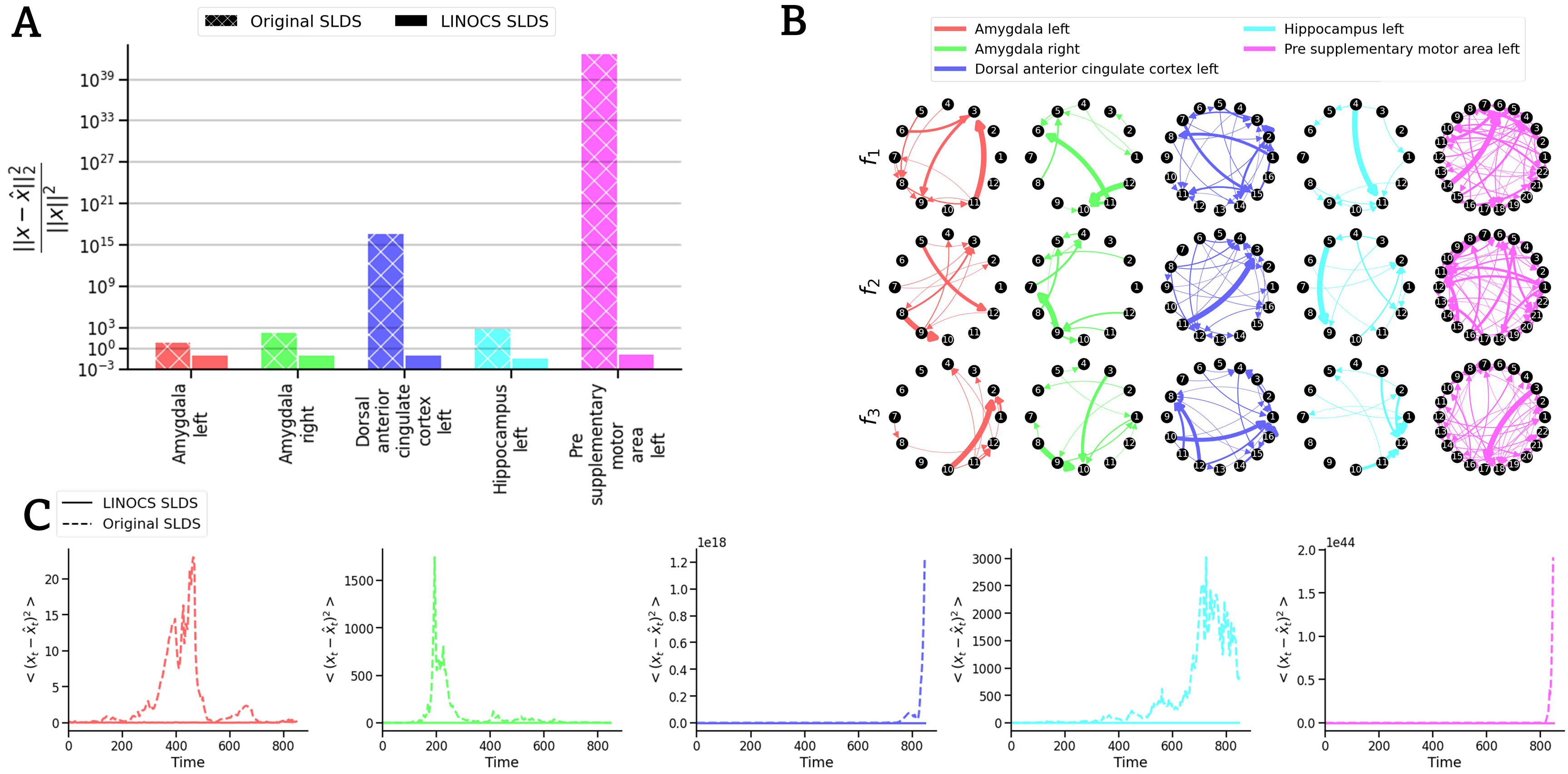}
    \caption{\textbf{SLDS results on real data.}
    \textbf{A:} Relative error of the full lookahead prediction compared to the ground truth, for both LINOCS-SLDS vs. classical SLDS,for each brain area.
    \textbf{B:}
    The networks identified by LINOCS-SLDS (see Fig.~\ref{fig:non_LINOCS_real_data_SLDS_operators} for the networks identified by the classical-SLDS).
    \textbf{C:} Lookahead reconstruction using LINOCS-SLDS (full curve) vs. classical SLDS (dashed curve). Classical-SLDS diverge  to extreme values in the full lookahead reconstruction.
        }
    \label{fig:SLDS_REAL}
\end{figure}

We next applied LINOCS-SLDS with three discrete states and compared it with regular SLDS using the same number of iterations. LINOCS identified operators that exhibit slight differences compared to those found by classical SLDS (Fig.~\ref{fig:SLDS_REAL}B vs Fig.~\ref{fig:non_LINOCS_real_data_SLDS_operators}; Fig.~\ref{fig:switches_and_operators_real_world} A vs. B) as well as slightly different switching patterns between the two approaches (Fig.~\ref{fig:switches_and_operators_real_world} C,D). These operators resulted in significantly more robust lookahead predictions. 
Specifically, differences are evident in both connection presence, weights, and distribution among global operators. For example, in the ``Amygdala left'' region, both classical SLDS and LINOCS-driven SLDS identify a connection from neuron 10 to neuron 2 as part of $\bm{f}_3$, albeit with varying weights. Additionally, both methods identify connections from 5 to 3 (in $\bm{f}_1$ for classical SLDS and in $\bm{f}_2$ for LINOCS-SLDS) as well as from 6 to 3 (in $\bm{f}_2$ for classical SLDS and in $\bm{f}_1$ for LINOCS-SLDS), but with differing weights. Similar discrepancies are observed in other regions. Furthermore, LINOCS-SLDS and classical-SLDS each identify connections that the other overlooks; for instance, in the ``Amygdala right'' region, LINOCS-SLDS identifies a strong connection from 11 to 6, whereas classical SLDS does not. Conversely, classical SLDS identifies a connection from 3 to 6 (in $\bm{f}_3$), which LINOCS-SLDS does not recognize.

Importantly, the operators found by LINOCS enable full lookahead reconstruction without diverging, in contrast to regular classical SLDS that diverge to extreme values in full lookahead prediction (Fig.~\ref{fig:SLDS_REAL}C). Moreover, the reconstruction error for the full lookahead prediction was overall much smaller for LINOCS-SLDS compared to the classical SLDS (Fig.~\ref{fig:SLDS_REAL}A).
These observations suggest that if the real neural process follows switched dynamics, LINOCS may capture the underlying dynamics more effectively, as inferred from our analysis of the synthetic case.

\begin{figure}[t]
    \centering
    \includegraphics[width=0.99\textwidth]{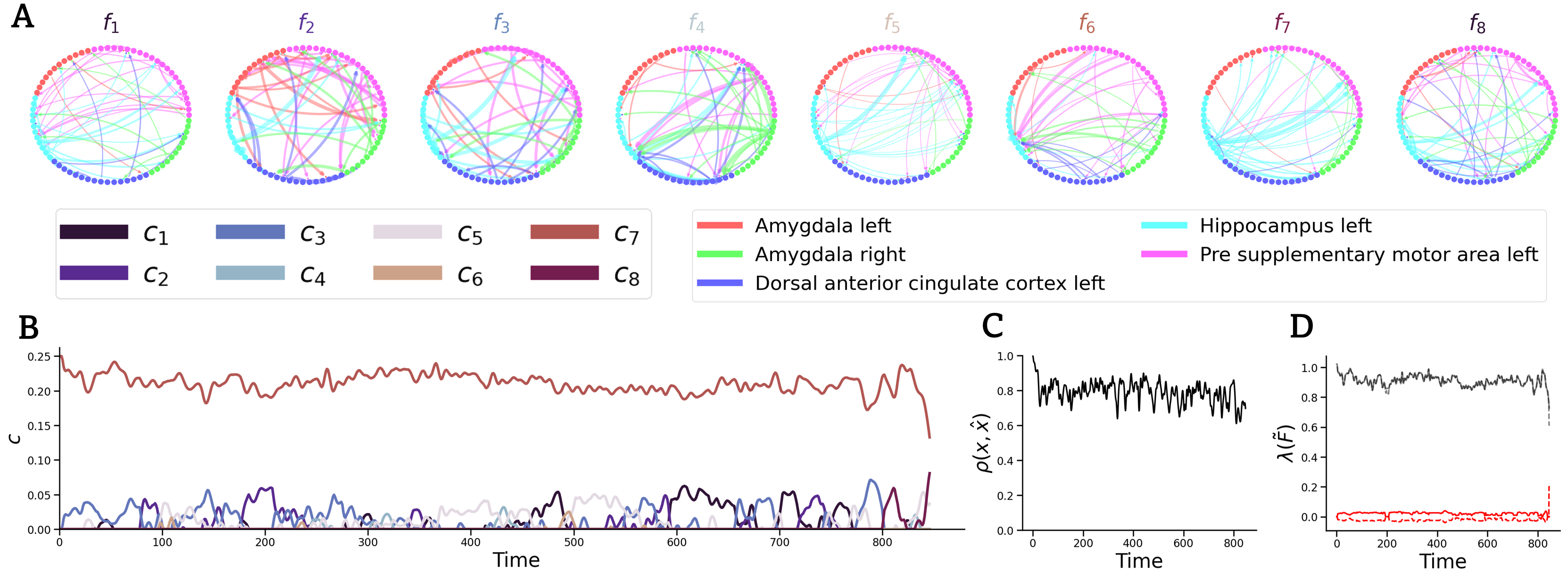}
    \caption{\textbf{dLDS-LINOCS results on the real neural data.}
    \textbf{A:} The identified operators $\{\bm{f}_j\}_{j=1}^{J=8}$.
    \textbf{B:}~The identified sparse coefficients $\bm{c}_t$.
    \textbf{C:} Correlation between the full-lookahead reconstruction results and the observations.
    \textbf{D:} The two largest eigenvalues of the time-varying transition operator $\widetilde{\bm{F}}_t = \sum_{j=1}^8 c_{jt}\bm{f}_j.$ Black: real part. Red: imaginary part. 
    }
    \label{fig:dlds_LINOCS_results_on_real_data}
\end{figure}

\begin{figure}[t]
    \centering
    \includegraphics[width=0.99\textwidth]{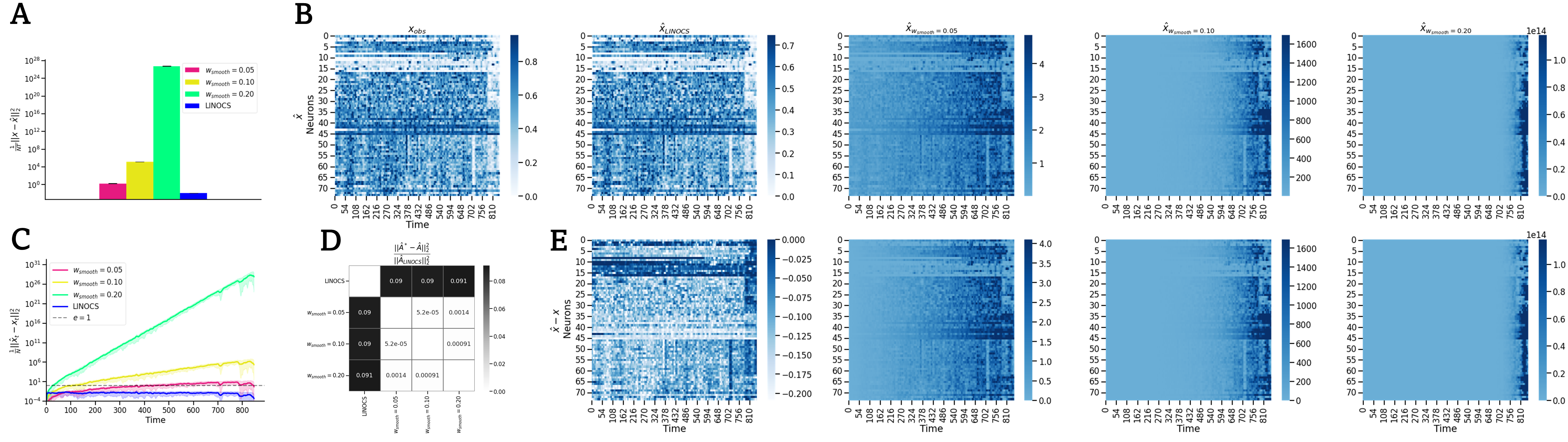}
    \caption{\textbf{LTV-LINOCS results on real neural data.}
    \textbf{A:} MSE between the ground truth data and the full lookahead reconstructions.
    \textbf{B:} The ground truth data (left) compared to the predictions produced by LINOCS-LTV (2nd left) and 1-step optimization with increasing smoothness constraints.
    \textbf{C:} MSE of the reconstruction over time points, compared to 1-step optimization with different smoothness levels.
    \textbf{D:} Frobenius 
    norm of the differences between $\bm{A}$s identified by the different models, normalized by the magnitude of the operator identified by LINOCS. 
    \textbf{E:} Difference between the observations and the full lookahead reconstruction for LINOCS-LTV vs. 1-step optimization with increasing smoothness levels.
    }
    \label{fig:LTV_LINOCS_results_on_real_data}
\end{figure}

We observed similar patterns using dLDS-LINOCS, which revealed underlying global brain interactions potentially fundamental to brain function (Fig.~\ref{fig:dlds_LINOCS_results_on_real_data} A). When examining their dynamic activations ($\bm{c}_t$), we noted a ``background'' interaction consistently active, with slight modulations over time (Fig.~\ref{fig:dlds_LINOCS_results_on_real_data} B, brown), alongside gradually changing activities of other interactions (Fig.~\ref{fig:dlds_LINOCS_results_on_real_data} B, gray-blue-purple). Importantly, these results provided lookahead predictions that did not decay and maintained a high correlation with the observations (Fig.~\ref{fig:dlds_LINOCS_results_on_real_data} C).

Finally, employed the LTV-LINOCS on all neurons from all regions simultaneously while imposing a smoothness constraint on consecutive operators
 (with regularization of $\lambda = 0.1$ on $\| \bm{A}_t - \bm{A}_{t-1}\|_2^2$, Fig.~\ref{fig:LTV_LINOCS_results_on_real_data}).
Our findings reveal that LINOCS identifies operators capable of producing full lookahead reconstructions without divergence, closely approximating observed data. Comparative analysis against 1-step optimization with various smoothness levels (Fig.~\ref{fig:LTV_LINOCS_results_on_real_data}A,B,E) underscores the superiority of LINOCS in achieving reconstructions faithful to the data. Additionally, examination of error evolution over time suggests a monotonic increase in error for non-LINOCS approaches (Fig.~\ref{fig:LTV_LINOCS_results_on_real_data}C). Moreover, we observed notable discrepancies between the operators identified by by LINOCS  and the baselines (Fig.~\ref{fig:LTV_LINOCS_results_on_real_data}D).
These results highlight the efficacy of LTV-LINOCS in capturing complex temporal dynamics in real world data while maintaining data fidelity. 
Overall, we showed that in all these real-world neural versions, LINOCS was able to recover more robust descriptions of the dynamic evolution for the long run, which, based on our synthetic results, may imply that these are closer to the real unknown interactions.


\section{Discussion}

In this paper we introduced LINOCS (Lookahead Inference of Networked Operators for Continuous Stability), a learning procedure to improve stability and accuracy of dynamical system inference that leverages lookahead estimation. By iteratively integrating re-weighted multi-step reconstructions with additional constraints on the operators, LINOCS enables robust inference of networked operators in dynamical systems, even in the presence of noise and nonlinearity.

Our experimental results highlight LINOCS' effectiveness across various dynamical systems, including Linear Systems (LDSs), Switched Linear Systems (SLDS), decomposed LDSs (dLDS)~\citep{mudrik2024decomposed}, and Linear Time-Varying Systems (LTV) in both simulation and real-world neural data. LINOCS not only achieves more precise full lookahead reconstruction compared to baseline methods but also successfully retrieves ground truth operators in synthetic data, signifying its superior capacity to capture underlying systems. These findings suggest that LINOCS holds greater potential than alternative approaches for accurately identifying unknown hidden interactions also in real-world data, where the real underlying interactions are often obscured but pivotal for robust scientific interpretation.

Looking ahead, several promising avenues exist for future directions and extensions, including applying LINOCS to improve the reconstruction robustness of highly non-linear systems and integrate it to advance robust RNN training. Particularly, if incorporating LINOCS to deep networks,  the integration of multi-step reconstructions into the networks' training, may help mitigate issues such as vanishing or exploding gradients. 
Additionally, extending LINOCS to handle non-Gaussian noise could enhance its applicability to a wider range of real-world scenarios.


\newpage

\newpage
\bibliographystyle{plainnat}
\bibliography{main}

\newpage
\appendix
\section{Appendix} 
\subsection{calculation details for operators differences (Fig.~\ref{fig:LINEAR_EXP}C)}\label{sec:cal_details_linear_diff}
We computed the operator differences illustrated in Figure~\ref{fig:LINEAR_EXP}C using the expression:

\[
((\widehat{\bm{A}} - \bm{I}) \bm{x} - (\bm{A} - \bm{I}) \bm{x}) \times \text{factor}
\]

Here, $\widehat{\bm{A}}$ represents the operators identified by the different methods, $\bm{A}$ denotes the ground-truth operators, and "factor" is a scalar used for visualization purposes only (identical for all methods).

\subsection{Hyperparamters used in experiments}\label{sec:hyperprameters}


\begin{table}[h!]
    \centering
    \caption{Hyperparameter settings for DAD baseline in linear experiment}
\begin{tabular}{|p{4cm}|p{3cm}|p{7cm}|}
\hline
\textbf{Parameter} & \textbf{Value} & \textbf{Additional Info} \\
\hline
seed & 0 & random seed\\ \hline
T & 500 &  number of time points\\ \hline
$w_{\ell_2}$ & 0 & weight of $\ell_2$ regularization on dynamics \\ \hline
$w_{decay}$ & 1 & decay of regularization coefficient over iterations \\ \hline
$N_{iterations}$ & 100 & number of iterations\\ \hline
A\_init\_type & step & initialize $A$ with 1-step optimization\\ \hline
 reweight & False & whether to reweight the observations and the lookahead during training. \\
\hline
    \end{tabular}
\end{table}

\begin{table}[h!]
    \centering
        \caption{Hyperparameter settings for ``DAD reweigh'' baseline in linear experiment}
    \begin{tabular}{|p{4cm}|p{3cm}|p{7cm}|}
        \hline
        \textbf{Parameter} & \textbf{Value} & \textbf{Additional Info} \\
        \hline
        seed & 0 & random seed \\ \hline
        T & 500 & number of time points \\ \hline
        $w_{\ell_2}$ & 0 & weight of $\ell_2$ regularization on dynamics \\ \hline
        $w_{decay}$ & 1 & decay of regularization coefficient over iterations \\ \hline
        $N_{iterations}$ & 100 & number of iterations \\ \hline
        A\_init\_type & 'step' & initialization type for matrix A \\ \hline
        reweight & True & whether to reweight the observations and the lookahead during training. \\
        \hline
    \end{tabular}
\end{table}

\begin{table}[h!]
    \centering
        \caption{Hyperparameter settings for ``reweigh $\ell_2$'' baseline in linear experiment}
    \begin{tabular}{|p{4cm}|p{3cm}|p{7cm}|}
        \hline
        \textbf{Parameter} & \textbf{Value} & \textbf{Additional Info} \\
        \hline
        seed & 0 & random seed \\ \hline
        T & 500 & number of time points \\ \hline
        $w_{\ell_2}$ & 1 & weight of $\ell_2$ regularization on dynamics \\ \hline
        $w_{decay}$ & 1 & decay of regularization coefficient over iterations \\ \hline
        $N_{iterations}$ & 100 & number of iterations \\ \hline
        A\_init\_type & 'step' & initialization type for matrix A \\ \hline
        reweight & True & whether to reweight the observations and the lookahead during training. \\
        \hline
    \end{tabular}
\end{table}

\begin{table}[h]
    \centering
        \caption{Hyperparameter settings for LINOCS in linear experiment}
    \begin{tabular}{|p{4cm}|p{3cm}|p{7cm}|}
        \hline
        \textbf{Parameter} & \textbf{Value} & \textbf{Additional Info} \\
        \hline
        K & 80 & maximum training order \\
        \hline
        with offset & True & whether to look for an offset term $\bm{b}$ \\
        \hline
        cal\_offset & True &  whether to calculate offset \\
        \hline
        weights\_style & exponential & weight style for orders\\
        \hline
        $\sigma_w$ & exponential parameter for the weights & 0.01 \\ \hline
        infer\_b\_way & 'each' & approach to infer the onset. Based on each order. \\
        \hline
        $K_b$ & 20 & maximum lookahead training order for the offsets $\bm{b}$ \\
        \hline
        weights\_style\_b & 'exponential' & weights style for the offsets $\bm{b}$ \\
        \hline
    \end{tabular}
\end{table}


\begin{table}[h]
    \centering
        \caption{Hyperparameter settings for dLDS experiment}
\begin{tabular}{|p{4cm}|p{3cm}|p{7cm}|}
    \hline
    \textbf{Parameter} & \textbf{Value} & \textbf{Additional Info} \\
    \hline
    K & 50 & maximum lookahead order \\
    \hline

    additional\_update & True & whether to include an additional update step \\
    \hline
    $\ell_1$\_init & 0 & value of $\ell_1$ on the coefficients for the $1$-st iteration  \\
    \hline
    max\_iters & 200 & maximum number of iterations \\
    \hline
    $\|F\|_2$ & 0 & $\ell_2$ norm on the basis of dLDS dynamics \\
    \hline
    $freq_{update_{F}}$ & 5 & frequency of updating $\{ \bm{f}_j \}_{j=1}^J$ \\
    \hline
     ${\ell_2}_\textrm{decay}$ & 0.99 & decay of the $\ell_2$ norm on the coefficients \\
    \hline
     ${\ell_1}_\textrm{decay}$ & 0.9999, &  decay of the $\ell_1$ norm on the coefficients \\ 
    \hline
     ${w_{smooth}}_\textrm{decay}$ & 1.01, & decay of smoothing weight on coefficients \\ \hline
     l\_smooth\_time & 1.1, &  regularization weight on coefficients smoothing $\bm{c}_t$ \\ \hline
    
     $\ell_2$ & 0, & $\ell_2$ on the coefficients \\ \hline

      $w_{smooth_{time}}$ & 0.1, & smoothness on $\bm{c}_t$ over time \\ \hline
     $\sigma_{noisy_c}$ & 0.05 & std of the noise to add to coefficients during training \\ \hline
   
     $\ell_1$ & 2.5, & $\ell_1$ regularization on $\bm{c}_t$ \\ \hline

    decor & False & wether to decorrelate the basis dynamics \\
    \hline
    max\_interval\_k & 1 & maximum interval to increase $k$ during training \\
    \hline
    to\_norm\_F & True & whether to normalize the basis dynamics \\
    \hline
    $\bm{x}_0$ & [0.2160895, 0.97627445, 0.00623026] & ground truth initial state (at $t=0$)\\
    \hline
    J & 3 & number of basis dynamics operators ($\bm{f}$s) \\
    \hline
    \end{tabular}
\end{table}

\begin{table}[h]
    \centering
        \caption{Hyperparameter setting for the LTV-Lorenz experiment}
    \begin{tabular}{|p{4cm}|p{3cm}|p{7cm}|}
    \hline
    \textbf{Parameter} & \textbf{Value} & \textbf{Additional Info} \\
    \hline
    $K$ & 5 & maximum lookahead order \\
       \hline
    $w_{smooth}$ & 2 (or 20) & smoothness weight \\
      \hline
    $estimate_{thres}$ & 2 & threshold to increase $k$ value \\
       \hline
    with\_future\_cost\_or & True & consider both fast and future reconstructions \\
       \hline
    $with_{future}$ & True & wether to apply time smoothness to future weights \\
       \hline
    $w_{smooth_{future}}$ & 0.05 & weights future smoothness \\
       \hline
    $weights_{style}$ & 'uni' & Uniform weight style. \\
       \hline
    $\|w\|_F$ & 0 & Frobenius norm weights \\
       \hline
    init\_style & 'step' &initialize based on optimal 1-step\\   \hline
    $max_{iters}$ & 40 & maximum number of iterations \\
       \hline
    $error_{thres}$ & 8 & threshold to increase error \\
       \hline
    $with_{reverse}$ & False & whether to include reverse update \\    \hline
    $N_{zeros}$ & 0 or 1 or 2 or 3 & how many zeros (relevant only to the sparse networks) \\
        \hline
    
    \end{tabular}
\end{table}

\subsection{Lorenz equations}\label{sec:lorenz}
The Lorenz attractor follows:
\begin{align}
\frac{dx}{dt} &= \sigma(y - x), \nonumber \\ 
\frac{dy}{dt} &= x(\rho - z) - y, \nonumber \\
\frac{dz}{dt} &= xy - \beta z, \nonumber
\end{align}
where \( x \), \( y \), and \( z \) represent the state variables, and \( \sigma \), \( \rho \), and \( \beta \) are the system parameters, set to \( \sigma = 10 \), \( \rho = 28 \), and \( \beta = \frac{8}{3} \).


\begin{figure}[t]
    \centering
    \includegraphics[width=0.9\textwidth]{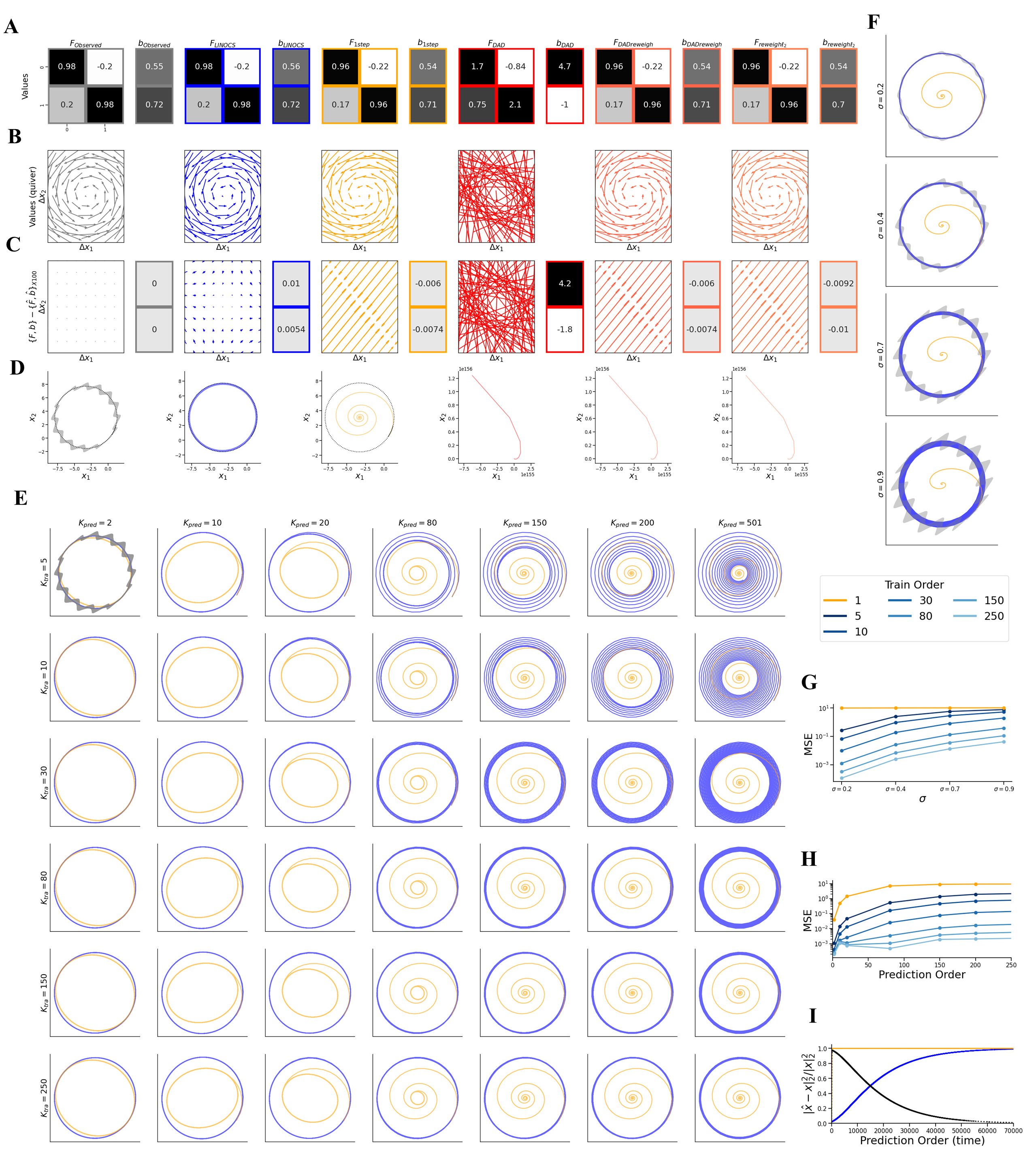} 
\caption{\textbf{Linear System under Structured Noise.} 
    \textbf{A:} Real vs. identified operators and offsets.
\textbf{B:} Quiver plots of real and identified operators present patterns that appear similar, rendering it challenging to discern differences when examined in isolation.
\textbf{C:} The differences in effects between real operators and inferred operators highlight how minor distinctions in dynamic operators gain prominence during lookahead reconstruction (calculation details in~\ref{sec:cal_details_linear_diff}).
    \textbf{D:} Full lookahead reconstruction (ground truth vs. baselines) shows swift convergence to the circle's center for the one-step optimization results due to small differences in dynamic values (mid-yellow subplot) and divergence for DAD-based results (three most-right subplots). 
    \textbf{E, H:} MSE under increasing prediction orders. LINOCS achieves better (lower) MSE compared to 1-step optimization. 
    \textbf{F, G:} LINOCS reconstruction compared to 1-step optimization under increasing noise values reveals that LINOCS maintains good reconstruction even under extreme noise conditions.
    \textbf{I:} By propagating identified operators until a relative reconstruction error of $\sim$ 1, LINOCS enables future predictions of $\sim$ 70,000 time points, contrasting with immediate convergence in one-step optimization. Black indicates error differences between one-step optimization and LINOCS.  
    }
    \label{fig:LINEAR_EXP_noise}
\end{figure}

\begin{figure}[h!]
    \centering
    \includegraphics[width=0.99\textwidth]{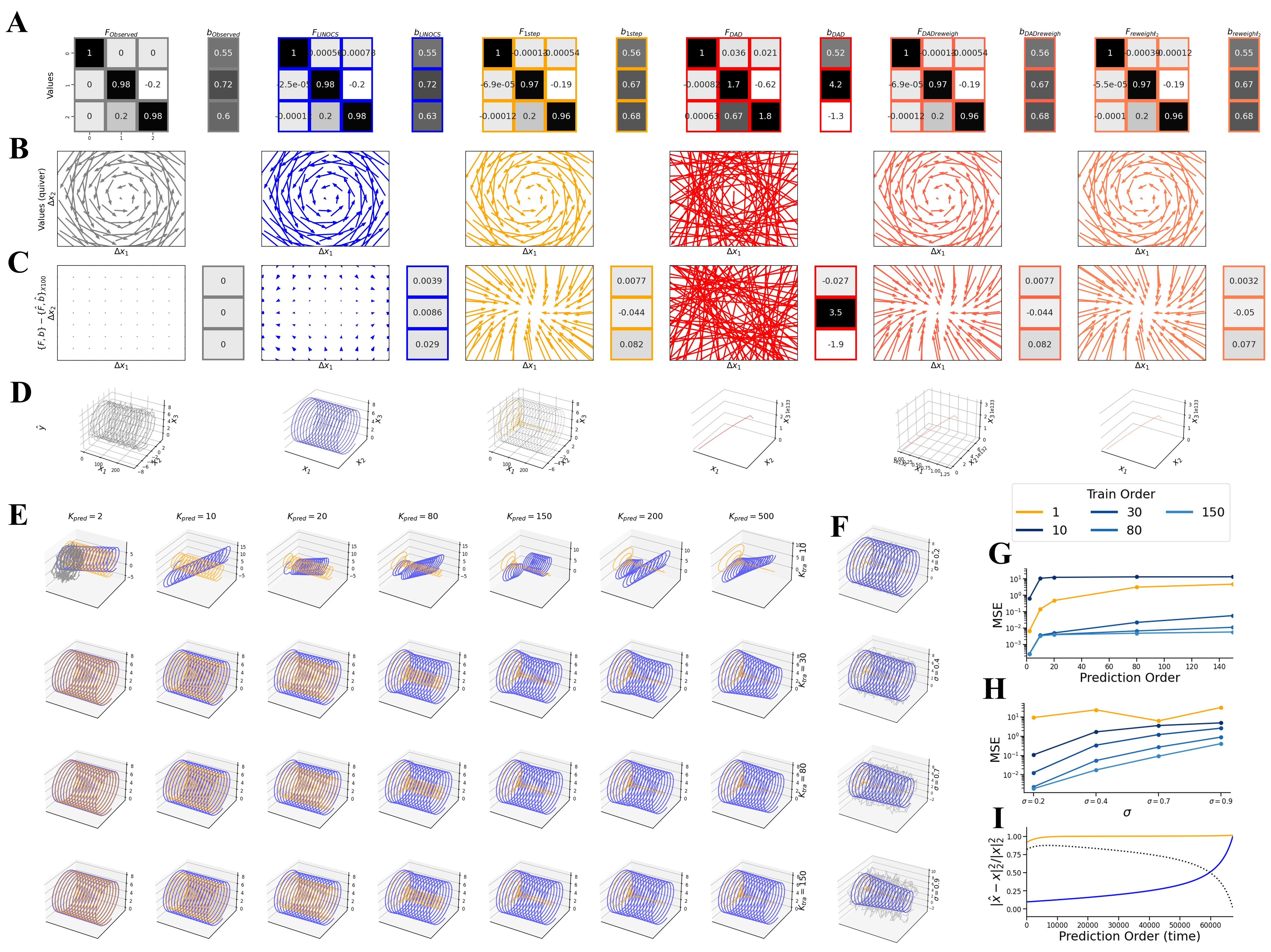}
\caption{\textbf{Linear System for $3D$ Cylinder.} 
    \textbf{A:} Real vs. identified operators and offsets.
\textbf{B:}~Quiver plots of real and identified operators present patterns that appear similar, rendering it challenging to discern differences when examined in isolation.
\textbf{C:} The differences in effects between real operators and inferred operators highlight how minor distinctions in dynamic operators gain prominence during lookahead reconstruction (calculation details in~\ref{sec:cal_details_linear_diff}).
    \textbf{D:} Full lookahead reconstruction (ground truth operators vs. baselines) shows swift convergence to the cylinder's center for the 1-step optimization results due to small differences in dynamic values (yellow) and divergence for DAD-based results (three most-right subplots). 
    \textbf{E, G:} MSE under increasing prediction orders. LINOCS achieves better (lower) MSE compared to 1-step optimization with perfect full lookahead reconstruction under high-enough training order (\textbf{E} right bottom). 
    \textbf{F,H:}~LINOCS reconstruction compared to 1-step optimization under increasing noise values reveals that LINOCS maintains good reconstruction even under extreme noise conditions.
    \textbf{I:}~Propagating the identified operators until reaching a relative reconstruction error of $\sim$ 1 shows that LINOCS identifies operators that enable a future prediction of $\sim$ 70,000 time points before converging, unlike one-step optimization that converges immediately. black: error difference between one-step optimization and LINOCS.    
    }
    \label{fig:LINEAR_EXP_3d}
\end{figure}


\begin{figure}[h]
    \centering
    \includegraphics[width=0.99\textwidth]{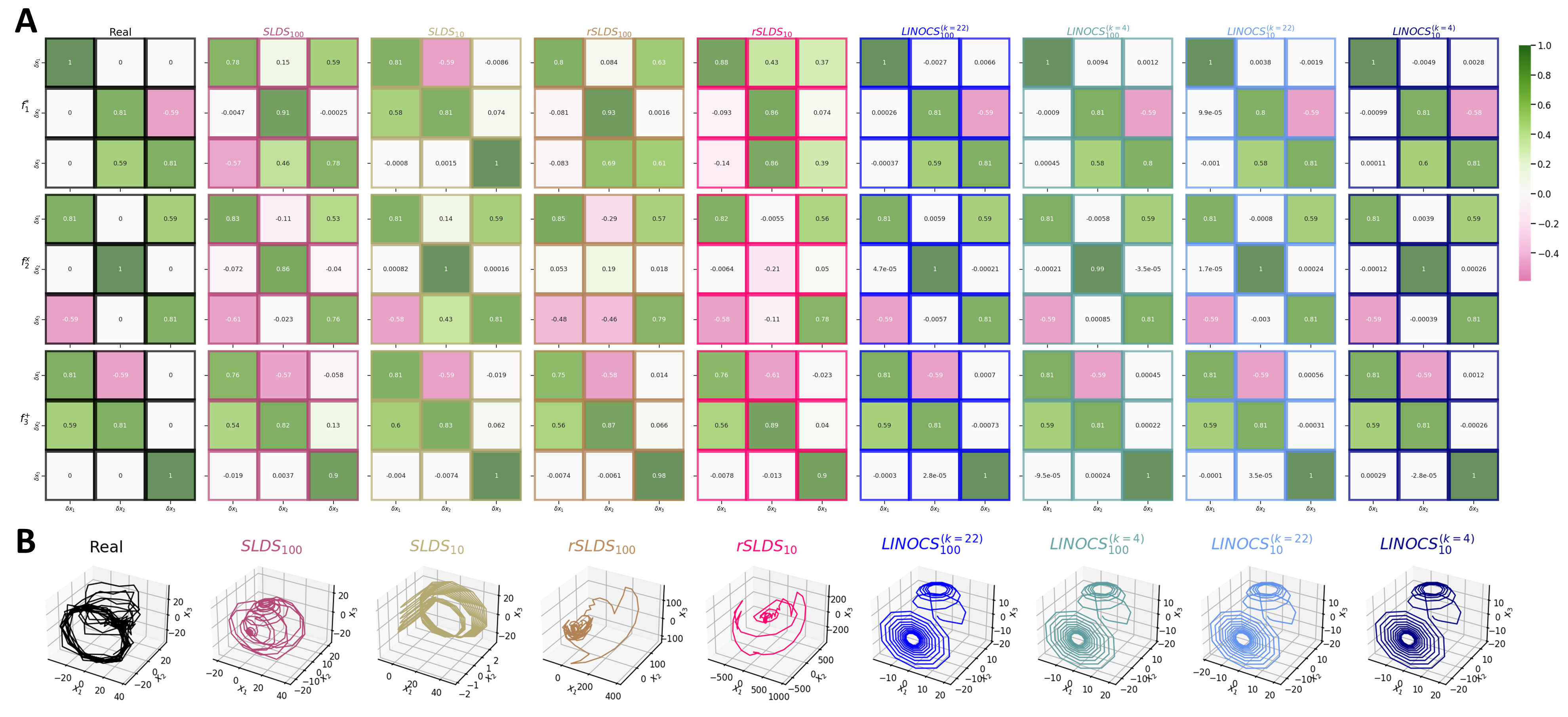}
    \caption{\textbf{Operators and reconstructions by the  LINOCS-driven SLDS compared to classical SLDS.} 
    \textbf{A:} Dynamical operators identified by LINOCS-driven SLDS.
    \textbf{B:} Full lookahead reconstruction.
    }
    \label{fig:switching_systems_supp}
\end{figure}

\begin{figure}[h]
    \centering
    \includegraphics[width=0.99\textwidth]{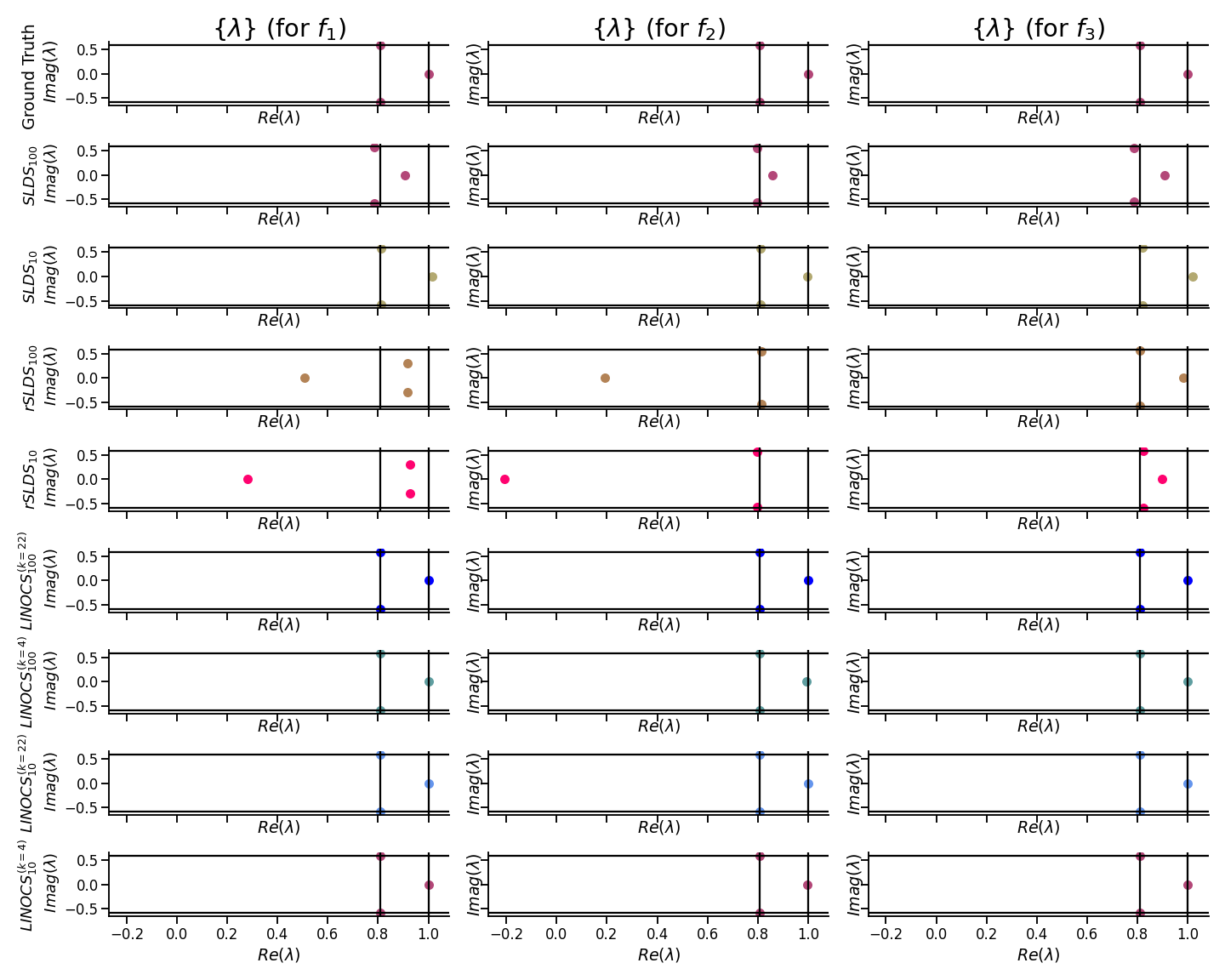}
   \caption{\textbf{Eigenvalues of identified operators by the different systems compared to the eigenvalues of the ground truth operators.} Rows represent different methods, with LINOCS (with different parameter combination) in the four last rows. Columns represent the three eigenvalues of each of the three different $3\times 3$ linear operators. LINOCS enabled the identification of almost perfect eigenvalues while the other methods found at least one wrong eigenvalue per operator, explaining the decaying/divergence of their reconstruction.
    }
    \label{fig:slds_eigenvalues}
\end{figure}

\begin{figure}[h]
    \centering
    \includegraphics[width=0.99\textwidth]{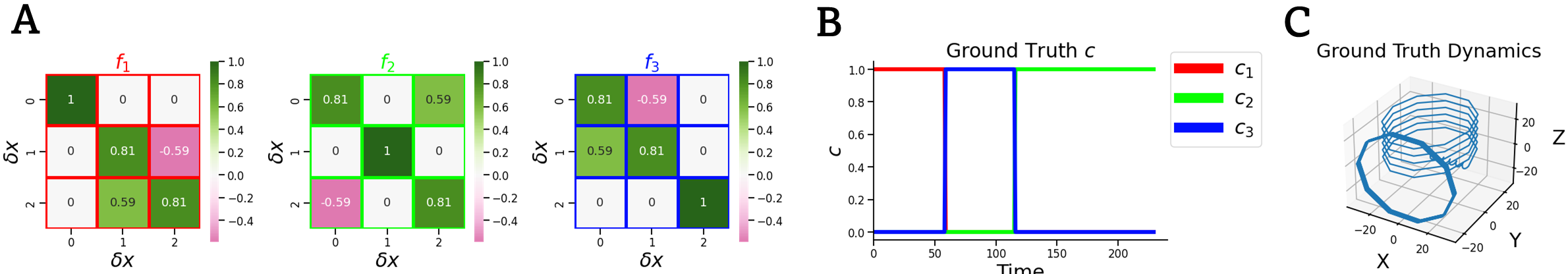}
 \caption{\textbf{Ground Truth operators and coefficients for the SLDS experiment.} 
\textbf{A:} The ground truth basis dynamics operators $\{\bm{f}_j \}_{j=1}^J$ consist of rotational matrices oriented in various directions.
    \textbf{B:}~Ground truth operators' coefficients ($\bm{c}$).
    \textbf{C:}  Ground truth state $\bm{x}$.
    }
    \label{fig:SLDS_synthetic_data}
\end{figure}

\begin{figure}[h]
    \centering
    \includegraphics[width=0.99\textwidth]{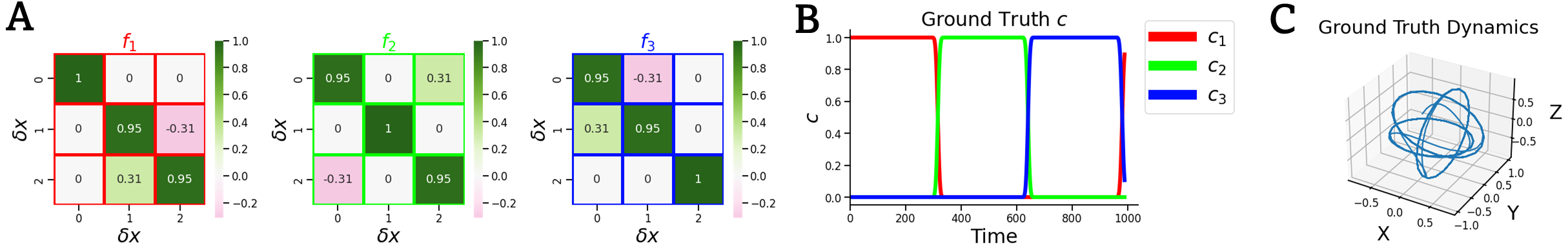}
    \caption{\textbf{Ground Truth operators and coefficients for the ``pseudo-switching'' dLDS experiment. } 
\textbf{A:} The ground truth basis dynamics operators $\{\bm{f}_j \}_{j=1}^J$ consist of rotational matrices oriented in various directions.
    \textbf{B:} Ground truth operators' coefficients ($\bm{c}$).
    \textbf{C:}  Ground truth state $\bm{x}$.
    }
    \label{fig:dlds_ground_truth}
\end{figure}

\begin{figure}[h]
    \centering
    \includegraphics[width=0.99\textwidth]{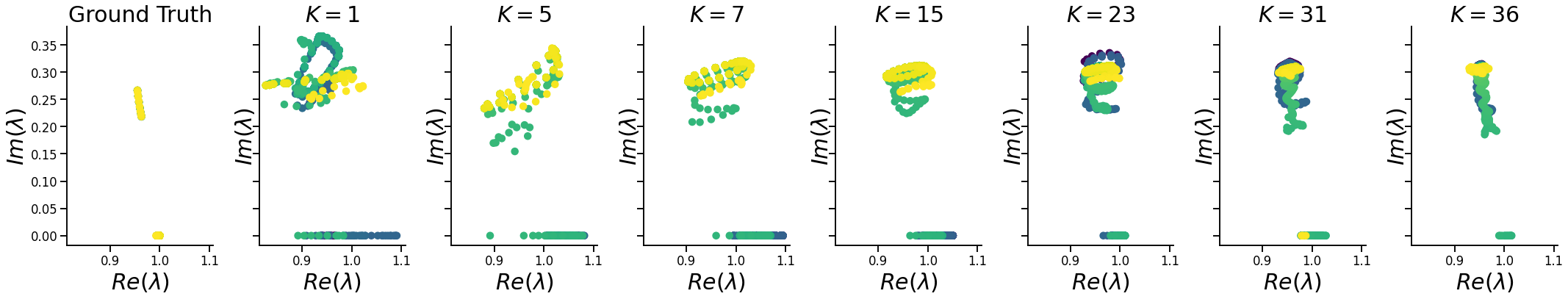}
    \caption{Real and imagery parts of the eigenvalues of the time-changing operator ${\bm{F}_t}$ captured by LINOCS-driven dLDS  }
    \label{fig:dlds_evals}
\end{figure}

\begin{figure}[h]
    \centering
    \includegraphics[width=0.99\textwidth]{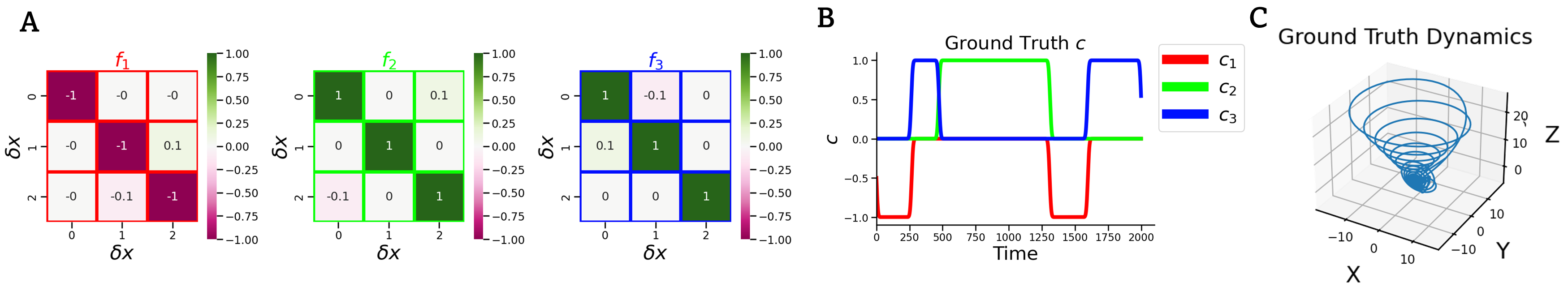}
    \caption{\textbf{Ground Truth operators and coefficients for the  second dLDS experiment.} 
\textbf{A:} The ground truth basis dynamics operators $\{\bm{f}_j \}_{j=1}^J$ consist of rotational matrices oriented in various directions.
    \textbf{B:} Ground truth operators' coefficients ($\bm{c}$).
    \textbf{C:}  Ground truth state $\bm{x}$.
    }
    \label{fig:dlds_ground_truth_complex}
\end{figure}



\begin{figure}[h]
    \centering
    \includegraphics[width=0.99\textwidth]{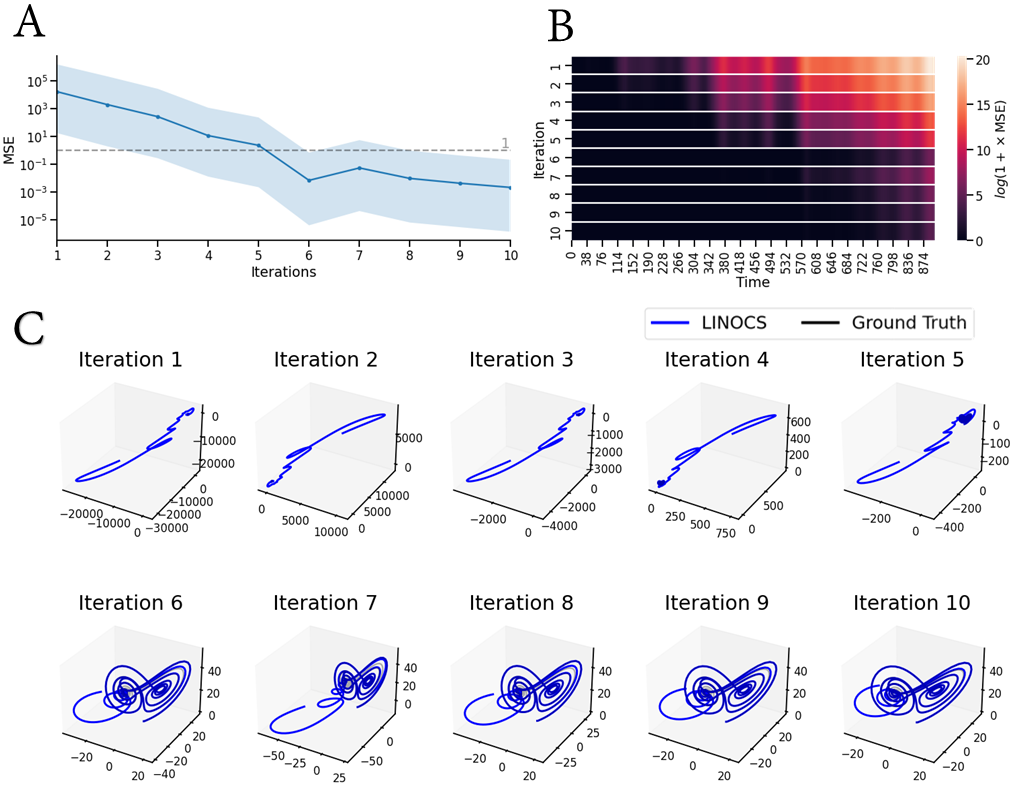}
    \caption{\textbf{Demonstration of LINOCS' effect over iterations for  Lorenz attractor with 900 time points.} 
    \textbf{A:} MSE over iterations (curve corresponds to median values; shade represents 25\%-75\% percentiles over time.
    \textbf{B:} MSE over time points over training iterations. 
    \textbf{C:}  Full lookahead reconstruction based on operators identified under different iterations.
    }
    \label{fig:LTV_ITERATIONS}
\end{figure}


\section{Appendix: Neural data additional information}

We further demonstrated LINOCS on multiple additional linear settings including data with structured noise (i.e., $\widetilde{\bm{x}} = \bm{x} + \sigma sin(3t)$, with $\sigma = 0.5$, Fig.~\ref{fig:LINEAR_EXP_noise}) and  a 3-dimensional cylinder (Fig.~\ref{fig:LINEAR_EXP_3d}). We found that LINOCS was able to  identify the ground truth dynamics for different training orders and under increasing noise levels also under these settings.

\begin{figure}[h]
    \centering    \includegraphics[width=0.99\textwidth]{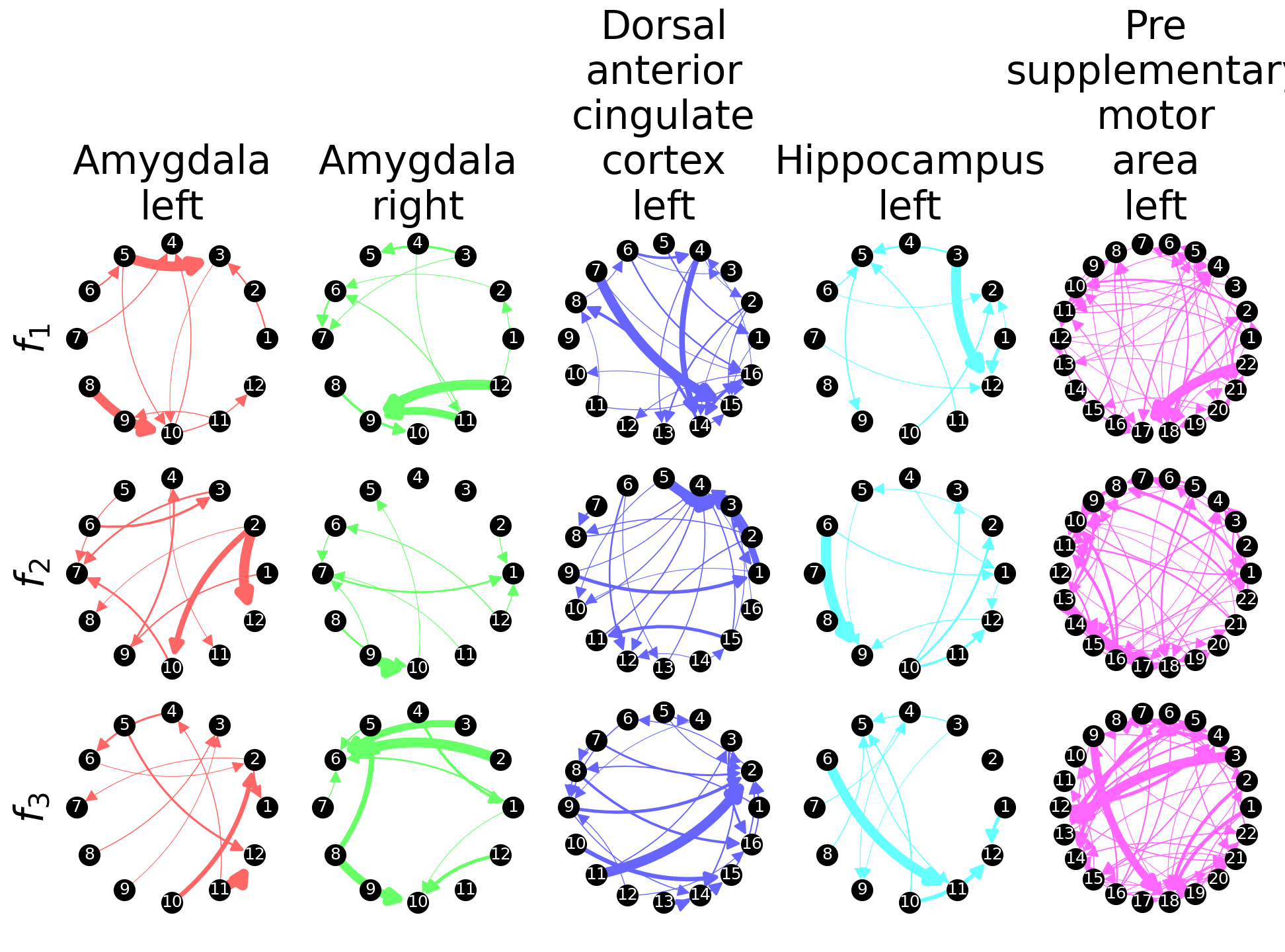}
   \caption{\textbf{Classical SLDS results (non LINOCS) on the real world data}. The identified networks ($\{ \bm{f}_j \}_{j=1}^J$) by the non-LINOCS SLDS code~\citep{Linderman_SSM_Bayesian_Learning_2020}, for each region in the real world data~\citep{kyzar2024dataset}. 
    }
    \label{fig:non_LINOCS_real_data_SLDS_operators}
\end{figure}


\begin{figure}[h]
    \centering
    \includegraphics[width=0.99\textwidth]{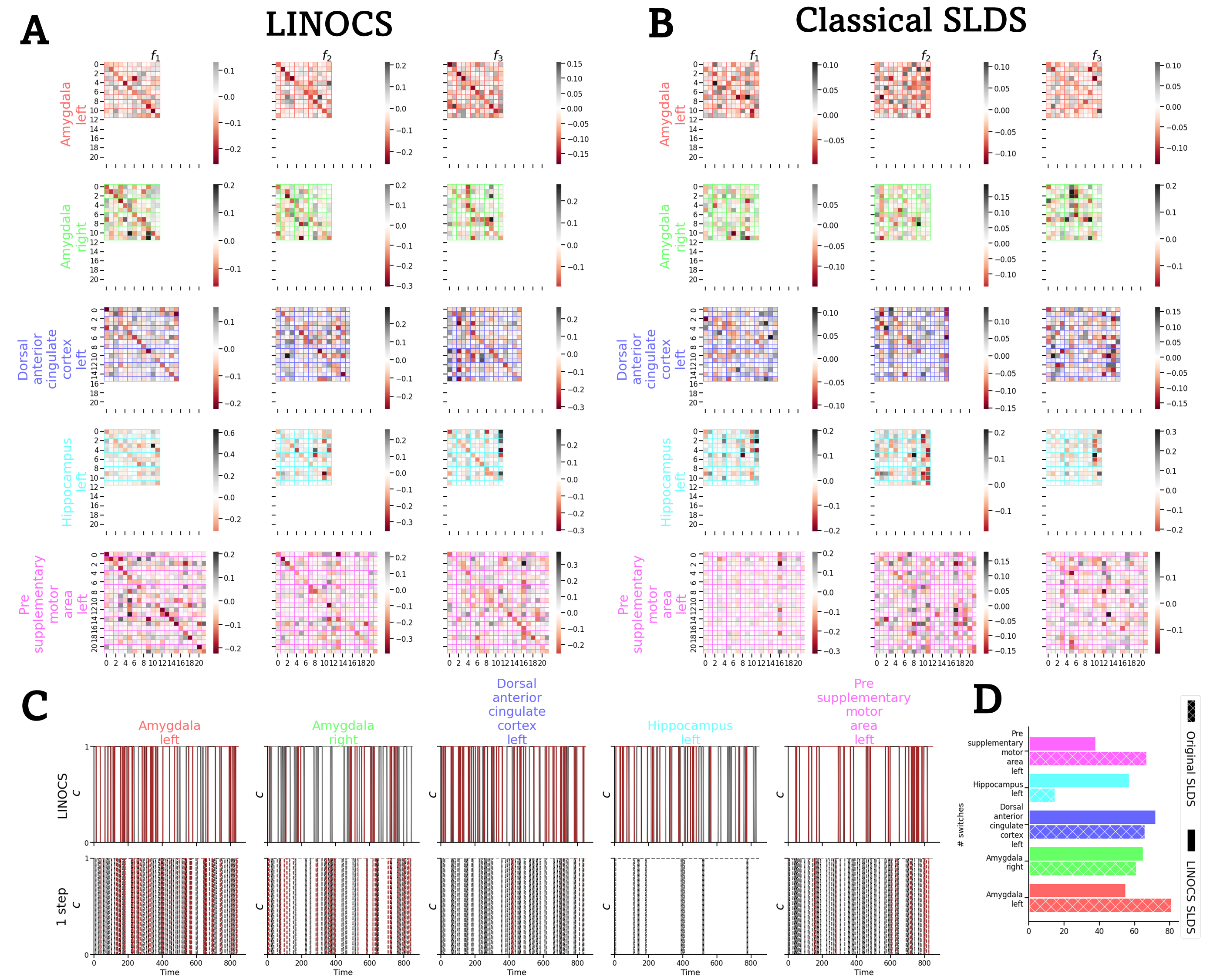}
    \caption{\textbf{SLDS results on real data.} 
    \textbf{A:}  Identified operators per region by LINOCS. 
    \textbf{B:} Identified operators per region by classical SLDS.  
    \textbf{C:} Switch times by LINOCS-SLDS vs classical SLDS. 
    \textbf{D:} Number of switches for LINOCS-SLDS vs. classical SLDS. 
    }
    \label{fig:switches_and_operators_real_world}
\end{figure}

\section{Code and Data Availability}
All algorithm implementation and figure creation codes will be shared over GitHub upon publication, and are attached as a supplement to the submission. The human neural recordings data is available at~\citep{kyzar2024dataset}.

\newpage

\end{document}